\documentclass[%
 reprint,
superscriptaddress,
showpacs,preprintnumbers,
 amsmath,amssymb,
 aps,
prb,
]{revtex4-1}

\usepackage{graphicx}
\usepackage{epstopdf}
\usepackage{dcolumn}
\usepackage{bm}
\usepackage{float}
\usepackage{hyperref}
\usepackage[caption=false]{subfig}
\usepackage{cancel}
\usepackage{afterpage}
\usepackage{xcolor}
\hypersetup{
    colorlinks=true,
    linkcolor={red!50!black},
    citecolor={blue!50!black},
    urlcolor={blue!80!black}
}

\begin{document}


\title{Scattering of magnetostatic surface modes of ferromagnetic films by geometric defects}

\author{R.E. Arias}%

\affiliation{%
Departamento de F\'isica, CEDENNA, Facultad de Ciencias F\'isicas y Matem\'aticas, Universidad de Chile
}%

\date{\today}

\begin{abstract}
Magnonics, an emerging field of Magnetism, studies spin waves (SWs) in nano-structures, with an aim towards possible applications. As information may be eventually transmitted with efficiency stored in the phase and amplitude of spin waves, a topic of interest within Magnonics is the propagation of SW modes. Thus, understanding mechanisms that may influence SW propagation is of interest. Here the effect of localized surface geometric defects on magnetostatic surface modes propagation is studied in ferromagnetic films and semi-infinite media. Theoretical results are developed that allow to calculate the scattering of these surface or Damon-Eshbach (DE) modes. A Green-Extinction theorem is used to determine the scattering of incident surface modes, through the determination of phase shifts of associated modes that are symmetric and anti-symmetric under inversion in the same geometry with geometric defects. Choosing localized symmetric depressions as geometric defects, scattering transmission coefficients are determined that show perfect transmission at specific frequencies or wave-lengths, that we associate with resonances in the system. Interestingly the system shows the appearance of localized modes in the depression regions, with associated discrete frequencies immersed in the continuum spectrum of these surface DE modes. These localized modes have a short wave-length content, and appear similarly in semi-infinite surfaces with depressions. The latter indicates that these types of scattering effects should appear in all surfaces with roughness or more pronounced geometric defects.
 
\end{abstract}

\maketitle

\section{Introduction:}

A present area of interest,  that has been developed mainly in the last couple of decades, is Magnonics \cite{Kruglyak2010,Chumak2015}, which studies spin waves in nano-structures \cite{Akhiezer1968,Gurevich1996} with an aim towards possible technological applications. Among advantages of spin waves to transmit information are low consumption of energy, long transmission lengths, and operation at microwave frequencies with wavelengths at the nano-scale, i.e. compatible with nano-circuits. Within Magnonics a particular topic of interest is propagation of spin waves in waveguides, since the control of proper transmission of information coded either in the amplitude or phase of a spin wave is crucial for applications. The present study corresponds to the latter topic of interest, since it delves with the effect of geometric defects on the propagation of magnetostatic surface waves or Damon-Eshbach (DE) \cite{DE1960} waves. In particular we study, as an example of localized geometric defects,  the effect of  depressions in the scattering of DE modes in ferromagnetic thin films and a semi-infinite media. 

Magnetostatic surface or volume modes have been studied in ferromagnetic films for a long time since the 1950's and even up to the present, in different ways. One may mention theoretical studies \cite{Patton1995,Kamenetskii2006}, and experimental ones \cite{Buttner2000,Bailleul2001,Wintz2019} that have dealt with these modes in ferromagnetic films.
Surface magnetostatic modes have the advantage that may propagate long distances, in the order of several microns \cite{Bailleul2001,Kostylev2014}, they may have large group velocities, and do not require large applied magnetic fields. 

The scattering of spin waves by different types of defects in films and wave-guides has been studied theoretically and experimentally. To mention a few studies, scattering by magnetic field non-uniformities was studied in Refs. \onlinecite{Slavin2004,Stamps2007}, by a localized defect in Ref. \onlinecite{Demidov2009}, by nano-defects and nano-wells \cite{Nafa2013,Krawczyk2019}, and by one dimensional steps \cite{Kostylev2019}. 

The surface geometric defects considered in the present theoretical study correspond to effectively 2D geometric features (no variation along a transverse direction) that do not alter the equilibrium magnetization of this Damon-Eshbach geometry: there is an in plane applied magnetic field with which the equilibrium magnetization aligns itself, and the surface spin waves propagate perpendicularly to this latter direction. The scattering of the magnetostatic surface waves is analyzed in these films using the Green-Extinction theorem method: it has been used by the author and co-workers in previous works, either in the magnetostatic approximation \cite{Mills2004,Mills2005,Jarufe2012} (the latter reference studied periodic surface geometric defects, with an associated appearance of frequency band-gaps) or in the dipole-exchange approximation \cite{Arias2016,Armijo2019}. The latter study introduced ``auxiliary functions" as part of the Green-Extinction method, an approach that is also used in the present study. The mentioned method allows to write integral equations for the modes evaluated on the surfaces (that may have arbitrary shapes in principle) and their corresponding eigen-frequencies. Indeed, after the integral equations are solved on the surface of the sample one may obtain the shape of the modes everywhere in space, if desired. 

The theoretical analysis of the magnetostatic scattering results has many similarities with a simpler case study that has been well researched and where a great deal of analytical progress has been possible: this corresponds to 1D scattering of a quantum mechanical particle by a localized potential. The associated Schrodinger equation has been well studied and scattering theories developed \cite{Eberly1965,Formanek1976,Sprung1996,Bliokh2005,Boya2008}.

An analysis of the flow of energy in the scattering process of a surface magnetostatic wave by surface geometric defects allows to determine reflection and transmission coefficients \cite{Gupta1979,Kamenetskii2006}. 
Also, an analysis in terms of the Green-Extinction theorem allows to determine the phase shifts of eigen-mode solutions that are ``symmetric" and ``anti-symmetric" under inversion in the presence of these geometric defects. As well as in the simpler 1D quantum particles scattering \cite{Eberly1965,Formanek1976}, the scattering solution of an incident magnetostatic surface wave may be related with these ``symmetric" and ``anti-symmetric" modes, and indeed the reflection and transmission coefficients may be directly written in terms of the difference between the previously mentioned phase shifts.  

Results are presented that apply the previous theory to the example of geometric defects that are depressions symmetrically located on both surfaces of the film. The transmission coefficient as function of frequency or incident wavelength presents resonances associated with perfect transmission, and also interestingly some localized modes appear at the location of the defects: the latter start appearing at short wavelengths, and they are also present with basically the same features in a semi-infinite medium.

\section{Magnetization dynamics}

 \subsection{Samples, magnetostatic surface modes configuration: \label{ssm}}
 
 We study scattering of magnetostatic surface modes in ferromagnetic films and semi-infinite media with geometric defects localized at the surfaces. We assume these defects to be geometric perturbations invariant in the $z$ direction,  and that there is a magnetic field applied in this direction, $\vec{H}_{app}=H_0 \hat{z}$: this determines that there is a uniform equilibrium magnetization, $M_s \hat{z}$, parallel to the applied magnetic field. We consider wave propagation in the $\pm \hat{x}$ directions with invariance along the transverse $z$ direction. Thus, the fields associated to the wave propagation vary effectively in two dimensions, the $x-y$ plane, with $y$ the direction perpendicular to the surfaces of the film or the semi-infinite medium. This geometry, magnetic field and wave propagation directions correspond to the so called Damon-Eshbach (DE) configuration, i.e. propagation perpendicular to the equilibrium magnetization, where it is well known that magnetostatic surface modes propagate \cite{DE1960}. 
  
 The perfect or un-perturbed geometries that we consider correspond to ferromagnetic films of thickness $w=2l$ (surfaces at $y=\pm l$), or a semi-infinite ferromagnetic medium at $y \geq 0$. 
 Given the applied magnetic field direction, in a semi-infinite medium these modes propagate only in one direction at frequency $\omega = |\gamma|(H_0+2\pi M_s)$ (in our case in the $\hat{x}$ direction; $\gamma$ is the gyromagnetic factor, $M_s$ the saturation magnetization). In a ferromagnetic film Damon-Eshbach surface modes propagating in different directions are reciprocal in frequencies but non-reciprocal in shape \cite{DE1960}: a right propagating mode ($\hat{x}$ direction) has its main amplitude associated with the lower surface of the film ($y=-l$), while a left propagating mode has amplitude mainly in the opposite upper surface. The dispersion relation of the magnetostatic DE modes is known analytically as \cite{DE1960}:
 \begin{equation}
 \Omega = \sqrt{(h_0+1/2)^2-e^{-4|k|l}/4}
  \label{DEf}
  \end{equation}
where $\Omega \equiv \omega/4\pi M_s|\gamma|$ represents normalized frequencies, $h_0 \equiv H_o/4\pi M_s$ a non dimensional magnitude of the applied magnetic field, and $k$ the wavevector of the surface modes. Thus, the lower end frequencies of this dispersion relation correspond to the long wavelength DE modes that have frequencies starting at $\Omega=\sqrt{h_0(h_0+1)}$ (which is the upper limit of the bulk modes), and the upper frequencies limit corresponds to the short wavelength surface modes that have frequencies that end at $\Omega=h_0+1/2$, i.e. at the mentioned frequency of surface waves in a semi-infinite medium. One may say that the finite thickness $w=2l$ of the film, or effectively the presence of two opposing surfaces, has opened up the degeneracy of the surface modes of a semi-infinite medium at the frequency $\Omega=h_0+1/2$. Indeed, the long wavelength surface modes change the most their frequencies, this may be understood since the amplitude of the surface modes penetrates a distance of the order of their wavelength $\lambda$ into the medium, i.e. the long wavelength modes “feel the effect" of the other surface of the film when $\lambda \sim l$, which is reflected in the dispersion relation of Eq. (\ref{DEf}). 
  
 The geometric defects that we consider alter the surfaces of the semi-infinite medium and the film, such that their new surfaces are described as $y=\xi (x)$ for the first, and as $y=l+\eta(x)$ for the upper surface of the film and $y=-l+\xi(x)$ for the lower. We consider these defects to be localized perturbations, i.e. the geometries are flat at $x \rightarrow \pm \infty$, and that they are even with respect to $x=0$ (this allows to simplify the analysis of the scattering problem in terms of spin wave modes with symmetry properties). Given these defects, the equilibrium magnetization is unaltered, it continues to be uniform and parallel to the geometric defects directions since these defects do not induce effective magnetic charges for $M_s \hat{z}$. 
 
 \subsection{Linear spin wave modes: \label{lsw}}
 
 The magnetization to linear order in these media may be written as:
 \begin{equation}
 \vec{M}(\vec{x},t) \simeq M_s \hat{z}+\vec{m}(\vec{x},t)  \; ,
 \label{}
 \end{equation}
 with $\vec{m}(\vec{x},t)=m_x \hat{x}+m_y \hat{y}$, i.e. perpendicular to $\hat{z}$, the equilibrium magnetization direction. Under these conditions we will determine linear surface eigenmodes of frequency $\omega$, as follows:
 
 \begin{equation}
 \vec{m}(\vec{x},t) = Re[\vec{m}^{\omega}(x,y) e^{-i \omega t}]  \; .
 \label{mod}
 \end{equation}
 
\subsubsection{Landau-Lifshitz equation:}

We are considering a micro-magnetic continuum model of description of the magnetization dynamics, which is governed by the Landau-Lifshitz equation of motion for the magnetization. In the magnetostatic approximation that we are considering the effective field that exerts torque on the magnetization is given by the sum of the applied magnetic field $H_0 \hat{z}$ and by the demagnetizing field $\vec{h}_D(\vec{m})$ produced by the dynamic magnetization. 
Then, the linear spin wave modes of Eq. (\ref{mod}) satisfy the following Landau-Lifshitz equation written to linear order:

\begin{equation}
i (\omega /|\gamma|) \vec{m} = (M_s \hat{z}+\vec{m}) \times (H_0 \hat{z}+\vec{h}_D) \; .
\label{LL}
\end{equation}
This leads to a linear relation between the components of the demagnetizing field and the dynamic magnetization:
\begin{equation}
\left( 
\begin{array}{c} 
h_D^x \\
h^y_D
\end{array}
\right) =
\left(
\begin{array}{cc} 
h_0 & i \Omega \\
-i \Omega & h_0
\end{array}
\right)
\left( 
\begin{array}{c} 
4 \pi m_x \\
4 \pi m_y
\end{array}
\right)
\; .
\end{equation}
 Inverting the previous relations, one obtains the following relation between the components of the dynamic magnetic induction $\vec{b}=\vec{h}_D+4\pi \vec{m}$ and those of the demagnetizing field:

\begin{equation}
\left( 
\begin{array}{c} 
b_x \\
b_y
\end{array}
\right) =
\left(
\begin{array}{cc} 
\mu & i \nu \\
-i \nu & \mu
\end{array}
\right)
\left( 
\begin{array}{c} 
h_D^x \\
h^y_D
\end{array}
\right)
\; ,
\label{bxh}
\end{equation}
with $\mu \equiv (h_0^2+h_0-\Omega^2)/(h_0^2-\Omega^2)$, $\nu \equiv -\Omega/(h_0^2-\Omega^2)$ frequency dependent effective susceptibility coefficients. These satisfy:
\begin{equation}
\mu-1 \pm \nu=1/(h_0 \pm \Omega) \; .
\label{eqsu}
\end{equation}

\subsubsection{Magnetostatic Maxwell equations, boundary conditions:}

The magnetic induction $\vec{b}$ and demagnetizing field $\vec{h}_D$ that the linear spin wave modes produce should satisfy the following Maxwell equations in the magnetostatic approximation:
\begin{equation}
\begin{array}{ccc}
\nabla \cdot \vec{b}=0 & , & \nabla \times \vec{h}_D=0
\end{array}
\label{mag}
\end{equation}
The second equation may be solved by introducing a magnetostatic potential $\phi (\vec{x},t)$ through $\vec{h}_D=-\nabla \phi$, and then the first becomes Laplace's equation for the magnetostatic potential both inside and outside the sample (since $\vec{b}=\vec{h}_D$ outside, and by use of Eqns. (\ref{bxh}) inside). The magnetostatic boundary conditions that these fields should satisfy on the surfaces of the ferromagnetic samples are that the normal component of the magnetic induction be continuous, i.e. $b_n=\vec{b} \cdot \hat{n}$, with $\hat{n}$ the normal to the sample surface, and that the tangential demagnetizing field be continuous, or equivalently that the magnetostatic potential be continuous.

\section{Green-extinction equations: \label{ge}}
 
 Instead of using the standard procedure for solving for the magnetostatic linear spin wave modes, that was explained in the previous section \ref{lsw},
the frequencies of the magnetostatic spin wave modes as well as their amplitudes on the surfaces of the sample may be obtained by solving integral equations satisfied by them
\cite{Mills2004,Mills2005}. These are homogeneous extinction equations that 
may be thought of 
as a generalization of Green's theorem to the equations relevant to this case, i.e. the magnetostatic equations for the demagnetizing fields and the Landau Lifshitz equation for the magnetization dynamics. In the following these
integral equations are derived for the magnetostatic normal modes evaluated on the surfaces of the sample.

 \subsection{Extinction equations outside the magnetized sample:}
 
 Extinction equations can be derived for the linear normal modes of the system if one integrates in the  region outside the magnetized sample the following expression
 that involves a so called “auxiliary function" \cite{Arias2016,Armijo2019}(in this case represented by $\phi_0^{-\omega}(\vec{x}-\vec{x}')$, with $\vec{x}'$ an arbitrary reference point). The latter satisfies the same magnetostatic equation in that region, as the modes, i.e. Laplace's equation, but it is evaluated at frequency $(-\omega)$. “Auxiliary" functions do not need to satisfy boundary conditions, and there lies their usefulness. The integrand is chosen as the following combination, and is null due to the first of the magnetostatic Eqns. (\ref{mag}):
 \begin{eqnarray}
0 & = &  \int_{V_{out}} dV \{ \phi_0^{-\omega}(\vec{x}-\vec{x}') \nabla \cdot \vec{b}^{\omega}(\vec{x})
\nonumber \\
& & 
-\phi^{\omega}(\vec{x}) \nabla \cdot \vec{b}_0^{-\omega}
(\vec{x}-\vec{x}') \}   \; .
 \end{eqnarray}
 The volume of integration is chosen outside the magnetized sample, but ending on its limiting surface. 
 Since outside $\vec{b}^{\omega} = - \nabla \phi^{\omega}$ and $\vec{b}_0^{-\omega} = - \nabla \phi_0^{-\omega}$,
 integrating by parts the previous equation one obtains:
  \begin{equation}
0= \int_S d\vec{S} \cdot \{ \phi_0^{-\omega}(\vec{x}-\vec{x}') \vec{b}^{\omega}(\vec{x})-\phi^{\omega}(\vec{x}) \vec{b}_0^{-\omega}(\vec{x}-\vec{x}') \} \; ,
 \label{ieo}
 \end{equation}
 with $S$ the surface of this non magnetized region, with its normal pointing outwards, but which effectively is the surface of the sample: one may change the sign, i.e. the normal may be taken pointing out of the sample (it is an homogeneous equation; the  surface at infinity does not contribute since the “auxiliary" function decays to zero at infinity).
 
 The “auxiliary" functions in the upper outside region may be taken as having the following simple form, characterized by a given wavevector $k$ and decaying at $y \rightarrow +\infty$: 
 \begin{eqnarray}
\phi_U^{-(\omega,k)} & = & e^{-i k x} e^{-|k| y} \\
b_U^y  & = & h_U^y= -\partial \phi_U/\partial y =|k| \phi_U \\
b_U^x & = &  h_U^x= -\partial \phi_U/\partial x = ik \phi_U \; ,
\end{eqnarray}
while those in the lower region are:
\begin{eqnarray}
\phi_L^{-(\omega,k)} & = & e^{-i k x} e^{|k| y} \\
b_L^y  & = & h_L^y= -\partial \phi_L/\partial y =-|k| \phi_L \\
b_L^x & = &  h_L^x= -\partial \phi_L/\partial x = ik \phi_L \; ,
\end{eqnarray}
their time depence is $\exp (i \omega t)$.
The extinction  Eq. (\ref{ieo}) in the upper region leads to (surface described by $y=l+\eta (x)$, $dl =dx \sqrt{1+\eta'(x)^2}$ the length differential, $\hat{n}=(-\eta'(x) \hat{x}+\hat{y})/\sqrt{1+\eta'(x)^2}$ the surface normal that points into the vacuum):
\begin{eqnarray}
0 & = & \int_{-\infty}^{\infty} dx e^{-ikx}e^{-|k| \eta (x)}
[\sqrt{1+(\eta')^2}b_n(x,l+\eta(x))
\nonumber \\
& & +
(ik \eta'(x)-|k|)\phi(x,l+\eta(x)) ] \; ,
 \label{exuf} 
\end{eqnarray}
and in the lower surface described by $y=-l+\xi (x)$ ($\hat{n}=(\xi'(x) \hat{x}-\hat{y})/\sqrt{1+\xi'(x)^2}$):
\begin{eqnarray}
0 & = &  \int_{-\infty}^{\infty} dx e^{-ikx}e^{|k| \xi (x)}
[\sqrt{1+(\xi')^2}b_n(x,-l+\xi(x))
\nonumber \\
& & -
(ik \xi'(x)+|k|)\phi(x,-l+\xi(x)) ] 
\label{exdf} 
\end{eqnarray}

\subsection{Extinction equations inside the magnetized sample: \label{exi}}
  
  In order to obtain extinction equations associated with the inside of a magnetized sample, we consider the following integral over a magnetized sample volume:
  \begin{equation}
 0 = \int_{V_{in}} dV \{ 
 \phi_I^{-\omega}(\vec{x},\vec{x}') \nabla \cdot \vec{b}^{\omega}  
 -\phi^{\omega}(\vec{x}) \nabla \cdot \vec{b}_I^{-\omega}(\vec{x},\vec{x}')
 \}  \; ,  \\
 \label{combG}
 \end{equation}
 with $\phi, \vec{b}$ corresponding to normal modes, and $\phi_I,\vec{b}_I$ representing “auxiliary" functions: both do satisfy the Landau Lifshitz  and magnetostatic equations, i.e. Eqns. (\ref{LL},\ref{mag}) (at frequency $(-\omega)$, $\vec{x}'$ a reference point, for the “auxiliary" function). 
 The integrand is null due to Maxwell's equation $\nabla \cdot \vec{b}=0$ (Eqns. (\ref{mag})).  Integrating by parts Eq. (\ref{combG}) and after some steps (they do require use of Eqns. (\ref{bxh})), one obtains:

 \begin{equation}
0 = \int_S d\vec{S} \cdot \{ 
\phi_I^{-\omega}(\vec{x},\vec{x}') \vec{b}^{\omega} (\vec{x})
-\phi^{\omega}(\vec{x}) \vec{b}_I^{-\omega}(\vec{x},\vec{x}')  \}
  \label{finex2}
 \end{equation}
 Inside the film the magnetostatic potential $\phi_I=\phi_{\pm}^{-(\omega,k)}$ of the “auxiliary" functions also satisfies Laplace's equation (it follows from Eqns. (\ref{bxh},\ref{mag})). We choose a pair of them, associated with a wavevector $(-k)$, and with growing and decaying exponential behaviors ($\pm$ signs) in the $y$ direction, as follows:
 
 \begin{eqnarray}
\phi_I=\phi_{\pm}^{-(\omega,k)} & = & e^{-i k x} e^{\pm|k| y} \; , 
\end{eqnarray}
\begin{equation}
\left( 
\begin{array}{c} 
b_x^{\pm}\\
b_y^{\pm}
\end{array}
\right) =
\left(
\begin{array}{cc} 
\mu & -i \nu \\
i \nu & \mu
\end{array}
\right)
\left( 
\begin{array}{c} 
h_x^{\pm}  \\
h_y^{\pm}
\end{array}
\right)
=
\left(
\begin{array}{c} 
i (\mu k \pm \nu |k|) \\
-(\nu k \pm \mu |k|) 
\end{array}
\right) \phi_{\pm}
\; .
\end{equation}

For the ($\pm$) “auxiliary" functions the previous extinction equations (\ref{finex2}) may be written as:
\begin{eqnarray}
0 & = & 
\int_{-\infty}^{\infty} dx e^{-i k x}e^{\pm |k| (l+\eta (x))} \{
\sqrt{1+(\eta'(x))^2}b_n(x,l+\eta(x))
 \nonumber \\ &  &+[(\nu k \pm \mu |k|)+i \eta'(x)(\mu k \pm \nu |k|)]\phi(x,l+\eta(x)) \} \nonumber \\ & + & 
\int_{-\infty}^{\infty} dx e^{-i k x}e^{\pm |k| (-l+\xi (x))} \{
\sqrt{1+(\xi'(x))^2}b_n(x,-l+\xi(x))
 \nonumber \\ &  &-[(\nu k \pm \mu |k|)+i \xi'(x)(\mu k \pm \nu |k|)]\phi(x,-l+\xi(x)) \}
\; .
\nonumber \\
\label{exif}
\end{eqnarray}

\subsection{Set of magnetostatic extinction equations in a film with geometric defects:}

In order to better handle the set of extinction equations (\ref{exuf},\ref{exdf},\ref{exif}) for the modes, we 
define $B_n^u(x) \equiv \sqrt{1+\eta'(x)^2}b_n(x,l+\eta(x))$ and $\Phi^u(x) \equiv \phi (x,l+\eta (x))$ in the upper surface, and similarly $B_n^d(x) \equiv \sqrt{1+\xi'(x)^2}b_n(x,-l+\xi(x))$ and $\Phi^d(x) \equiv \phi (x,-l+\xi (x))$ in the lower one. Thus, the Eqns. (\ref{exuf},\ref{exdf},\ref{exif}) become the following system of magnetostatic extinction equations:
\begin{eqnarray}
0 & = &  \int_{-\infty}^{\infty} dx e^{-ikx}e^{-|k| \eta (x)}
\{ B_n^u(x)+
(ik \eta'(x)-|k|)\Phi^u(x) \} \label{eu} 
\nonumber \\
0 & = &  \int_{-\infty}^{\infty} dx e^{-ikx}e^{|k| \xi (x)}
\{ B_n^d(x)-
(ik \xi'(x)+|k|)\Phi^d(x) \} \label{ed} 
\nonumber \\
0 & = & 
\int_{-\infty}^{\infty} dx e^{-i k x}e^{\pm |k| (l+\eta (x))} \{
B_n^u(x)
\nonumber \\ & & 
+[(\nu k \pm \mu |k|)+i \eta'(x)(\mu k \pm \nu |k|)]\Phi^u(x) \} \nonumber \\ & + & 
\int_{-\infty}^{\infty} dx e^{-i k x}e^{\pm |k| (-l+\xi (x))} \{
B_n^d(x)
\nonumber \\ & & 
-[(\nu k \pm \mu |k|)+i \xi'(x)(\mu k \pm \nu |k|)]\Phi^d(x) \}
\nonumber \\
\label{ef}
\end{eqnarray}

\subsection{Even geometric obstacles, using symmetry properties:}

If we consider even geometric obstacles, i.e. $\eta(-x)=\eta (x)$ and $\xi (-x)=\xi (x)$, one may analyze the spin wave modes in terms of functions with different symmetries with respect to the plane $x=0$, i.e. even or odd. For example the magnetostatic potential evaluated at the upper surface is separated into even (e) an odd (o) parts:
\begin{equation}
\Phi^u(x) = \Phi_e^u(x)+\Phi_o^u(x)
\end{equation}

It is convenient to  describe even, $E(x)$, and odd, $O(x)$, functions in terms of cosine, $E_c(q)$, and  sine, $O_s(q)$, Fourier transforms, as follows:
\begin{eqnarray}
E (x) & = & 
\frac{1}{\pi} \int_{0}^{\infty} dq \cos (qx) E_c (q) 
\label{icft} \nonumber \\ 
O (x) & = &
\frac{1}{\pi} \int_{0}^{\infty} dq \sin (qx) O_s (q) 
\label{isft} \nonumber \\
E_c(q) & = & = 2\int_{0}^{\infty} dx \cos (qx)  E (x) 
 \label{cft} \nonumber \\ 
 O_s(q) & = &  2 \int_{0}^{\infty} dx \sin (qx) O(x) \;  ,
\label{sft} \nonumber \\
\end{eqnarray}
with more details in Appendix \ref{scFT}.

Thus, the equations for the spin wave modes that follow from the Extinction  Eqns. (\ref{ef}) become (details in Appendix \ref{sam}, $s(k) \equiv sign(k)$):

\begin{eqnarray}
0 & = &  B_e^u(k) -is(k)B_o^u(k)-H_e^u(k) +is(k)H_o^u(k) 
\nonumber \\ & +&
\frac{2}{N} \sum_{q}  \{ C_u^{-|k|}(k,q) B_e^u(q) 
-i s(k) S_u^{-|k|}(k,q) B_o^u(q)  \nonumber \\ & - & 
S_u^{-|k|}(k,q) H_e^u(q)
+is(k)C_u^{-|k|}(k,q) H_o^u(q)
 \} \label{utd} 
 \\ 
 0 & = & B_e^d(k) -is(k)B_o^d(k) -H_e^d(k) 
 +is(k)H_o^d(k) 
 \nonumber \\ & +&
 \frac{2}{N} \sum_{q}  \{  C_d^{|k|}(k,q) B_e^d(q) 
 -i s(k) S_d^{|k|}(k,q) B_o^d(q) \nonumber \\ & -& 
S_d^{|k|}(k,q) H_e^d(q) +is(k)C_d^{|k|}(k,q) H_o^d(q) \}
 \label{dtd} 
  \\ 
0 & = & e^{\pm |k|l} \{
(s(k)\Omega \pm h) (B_e^u(k)-is(k)B_o^u(k) )
\nonumber \\ & + & 
(h \pm s(k)\Omega+1) (H_e^u(k)-is(k) H_o^u(k) )
 \nonumber \\ &+ & 
\frac{2 }{N}  \sum_{q} [ (s(k)\Omega \pm h) (C_u^{\pm|k|}(k,q) B_e^u(q)
 \nonumber \\ & - & 
 is(k) S_u^{\pm|k|}(k,q) B_o^u(q) )
+ (h \pm s(k) \Omega+1)
 \nonumber \\ & & 
( S_u^{\pm|k|}(k,q) H_e^u(q) 
-i s(k) C_u^{\pm|k|}(k,q) H_o^u(q) ) ] \}
\nonumber \\
& + & e^{\mp |k|l} \{
(s(k) \Omega \pm h) (B_e^d(k)-is(k)B_o^d(k) )
\nonumber \\ & & 
-(h \pm s(k) \Omega+1) (H_e^d(k) -is(k)H_o^d(k) )
\nonumber \\ & + & 
\frac{2}{N}  \sum_{q}  [ (s(k) \Omega \pm h) (C_d^{\pm|k|}(k,q) B_e^d(q)
 \nonumber \\ &- & 
is(k) S_d^{\pm|k|}(k,q) B_o^d(q) )  
-(h \pm s(k) \Omega+1) 
 \nonumber \\ & & 
(S_d^{\pm|k|}(k,q) H_e^d(q)
-is(k) C_d^{\pm|k|}(k,q) H_o^d(q) ) ] \} 
\nonumber \\ 
 \label{mtd} 
\end{eqnarray}

with:
\begin{eqnarray}
C_u^{\pm|k|}(k,q) & \equiv &   \int_{-\infty}^{\infty} dx \cos(qx)\cos(kx)(e^{\pm |k| \eta (x)}-1)
\nonumber \\
C_d^{\pm|k|}(k,q) & \equiv &   \int_{-\infty}^{\infty} dx \cos(qx)\cos(kx)(e^{\pm|k| \xi (x)}-1) \nonumber \\
S_u^{\pm|k|}(k,q)  & \equiv &   s(k) \int_{-\infty}^{\infty} dx \sin(qx)\sin(kx)(e^{\pm|k| \eta (x)}-1)
\nonumber \\
S_d^{\pm|k|}(k,q)  & \equiv &    s(k) \int_{-\infty}^{\infty} dx \sin(qx)\sin(kx)(e^{\pm|k| \xi (x)}-1)
\nonumber \\
\end{eqnarray}

The previous Extinction Eqns. (\ref{mtd}), were  written in terms of sine and cosine Fourier transforms coefficients (they are the unknowns, notice that they are even in their arguments, $k$ or $q$; also the eigenvalue $\Omega$ is an unknown). New variables $H_e(k) \equiv |k| \Phi_e (k)$, $H_o(k) \equiv |k| \Phi_o (k)$ were introduced, and we also used Eq. (\ref{eqsu}) that allows at the end to transform the Extinction equations into a standard matrix eigenvalue problem for the frequency $\Omega$. Furthermore, the integrals and transforms were written in their discrete versions, a necessary step for numerical calculations (see Appendix \ref{scFT}).

\subsection{ ``Mirror reflected" obstacles along the thickness direction of the film:}

We also consider ``mirror reflected" obstacles along the thickness direction $y$ (with respect to the plane $y=0$), which correspond to “symmetric" perturbations, i.e. $\eta (x)=-\xi(x)$. For them $C_u^{\pm |k|}=C_d^{\mp |k|} \equiv C_{\pm}$, and also $S_u^{\pm |k|}=S_d^{\mp |k|} \equiv S_{\pm}$. Also, we define:

\begin{eqnarray}
    P_{\pm}(k,q) & = & \delta_{k,q}+\frac{2}{N}C_{\pm}(k,q) \nonumber \\
    R_{\pm}(k,q) & = & \delta_{k,q}+\frac{2}{N}S_{\pm}(k,q)
    \nonumber \\ \label{pr} \\
    M_{\pm}(k,q) & = & e^{\pm |k|l}P_{\pm}(k,q)  \nonumber \\
    N_{\pm}(k,q) & = & e^{\pm |k|l}R_{\pm}(k,q)
    \label{mn}
\end{eqnarray}

Thus, for these ``mirror reflected" obstacles and taking the cases $s=sign(k)=\pm 1$, the previous Extinctions Eqns. (\ref{utd}-\ref{mtd})  may be written more compactly as:
\begin{eqnarray}
  0 & = & P_- B_e^u -R_- H_e^u \label{ue} \\
  0 & = & R_- B_o^u -P_- H_o^u \label{uo} \\ 
  0 & = &  P_- B_e^d -R_- H_e^d \label{de} \\
  0 & = & R_- B_o^d -P_- H_o^d \label{do}  \\
0 & = & (\Omega+h) [M_+ B_e^u+M_-B_e^d
 -iN_+ B_o^u-iN_-B_o^d) ]
 \nonumber \\ & +& (h+\Omega+1)[N_+ H_e^u-N_-H_e^d
 -iM_+ H_o^u+iM_-H_o^d ] \nonumber \\
 \label{e1}  \\
 0 & = &  -(\Omega+h)[M_- B_e^u+M_+B_e^d
 +iN_- B_o^u+iN_+B_o^d]
 \nonumber \\ &+ & (h+\Omega+1)[N_- H_e^u-N_+H_e^d
 +iM_- H_o^u-iM_+H_o^d]
 \nonumber \\ \label{e2}  \\
 0 & = & -(\Omega-h) [M_+ B_e^u+M_-B_e^d
 +iN_+ B_o^u+iN_-B_o^d) ]
 \nonumber \\ & +& (h-\Omega+1)[N_+ H_e^u-N_-H_e^d
 +iM_+ H_o^u-iM_-H_o^d ] \nonumber \\
 \label{e3}  \\
 0 & = &  (\Omega-h)[M_- B_e^u+M_+B_e^d
 -iN_- B_o^u-iN_+B_o^d]
 \nonumber \\ &+ & (h-\Omega+1)[N_- H_e^u-N_+H_e^d
 -iM_- H_o^u+iM_+H_o^d]
 \nonumber \\ \label{e4} 
\end{eqnarray}
We define symmetric and anti-symmetric variables with respect to the thickness direction $y$: $B_e^s=B_e^u+B_e^d$, $B_e^a=B_e^u-B_e^d$, and analogously for the variables $B_o$, $H_e$, $H_o$. Indeed, due to our assumed symmetries of the geometric defects, i.e. a geometry with mirror reflection  properties with respect to the planes $x=0$ and $y=0$, the modes separate into symmetric and anti-symmetric with respect to the inversion transformation $(x,y) \rightarrow -(x,y)$, and are associated with the variables $B_e^s, B_o^a, H_e^s, H_o^a$, and $B_o^s, B_e^a, H_o^s, H_e^a$ respectively. This happens, because given the applied magnetic field and the assumed form of the defects the system is invariant under the  inversion transformation $(x,y) \rightarrow -(x,y)$ or equivalently by a rotation of the system in $180$ degrees with respect to the $z$ axis. This separation of the problem into the just called symmetric and anti-symmetric modes is seen explicitly to happen as follows. 
Summing Eqns. (\ref{e1},\ref{e2}) and Eqns. (\ref{e3},\ref{e4}), and also from Eqns. (\ref{ue},\ref{de}) and Eqns, (\ref{uo},\ref{do}), one obtains the following equations:

\begin{eqnarray}
 0 & = & (\Omega+h) [(M_+ -M_-)B_e^a
 -i(N_+ +N_-) B_o^s ]
 \nonumber \\ & +& (h+\Omega+1)[(N_+ +N_-)H_e^a
 -i(M_+ -M_-)H_o^s ] \nonumber \\
 \label{a1}  \\
  0 & = & -(\Omega-h) [(M_+ -M_-)B_e^a
 +i(N_+ +N_-)B_o^s) ]
 \nonumber \\ & +& (h-\Omega+1)[(N_+ +N_-)H_e^a
 +i(M_+ -M_-)H_o^s ] \nonumber \\
 \label{a2}  \\
  0 & = & P_- B_e^a -R_- H_e^a \label{a3} \\
  0 & = & R_- B_o^s -P_- H_o^s \label{a4}  
 \end{eqnarray}
  The previous Eqns. (\ref{a1}-\ref{a4}) correspond to modes that are anti-symmetric with respect to the inversion  symmetry operation. 
Now, summing and subtracting Eqns. (\ref{a1},\ref{a2}) one obtains:
 \begin{eqnarray}
 0 & = & -\Omega (N_sB_o^s+M_dH_o^s)
 -ih(N_sH_e^a+M_dB_e^a)-iN_sH_e^a
 \nonumber \\ \label{af1} \\
 0 &= &   -\Omega (N_sH_e^a+M_dB_e^a)
 +ih (N_sB_o^s+M_dH_o^s) +iM_d H_o^s
  \nonumber \\ \label{af2}
  \end{eqnarray}
  with
  \begin{equation}
  \begin{array}{ccc}
   N_s \equiv N_+ +N_-$, $M_d \equiv M_+ -M_-
   \label{nsmd}
   \end{array}
   \end{equation}
   Defining new variables:
  \begin{equation}
  \begin{array}{ccc}
  N_sB_o^s+M_dH_o^s=WH_o^s
  & , & N_sH_e^a+M_dB_e^a=XH_e^a 
  \label{wxa}
  \end{array}
  \end{equation}
   with, from Eqns. (\ref{a3},\ref{a4}):
  \begin{equation}
  \begin{array}{ccc}
  W \equiv N_sV+M_d
  & , & X \equiv N_s+M_dT \\
  V \equiv R_-^{-1}P_- & , & T \equiv P_-^{-1}R_- \; ,
  \label{vta}
  \end{array}
  \end{equation}
then Eqns. (\ref{af1},\ref{af2}) may be written as:
    \begin{equation}
0=
\left(
\begin{array}{cc}
-\Omega I & -ih I-iN_sX^{-1} \\
iM_dW^{-1} + i h I& -\Omega I
\end{array}
\right) 
\left(
\begin{array}{c}
W H_o^s \\
X H_e^a
\end{array}
\right)
\label{eqa}
\end{equation}
  
  Notice that from Eqns. (\ref{mn}) the matrices $M_{\pm}, N_{\pm}$ may be written as:
\begin{equation}
\begin{array}{ccc}
 M_{\pm}=D_{\pm}P_{\pm} & , & N_{\pm}=D_{\pm}R_{\pm} \; ,
 \end{array}
 \label{mns}
 \end{equation}
 with $D_{\pm}$ the following diagonal matrices:
\begin{equation}
\begin{array}{ccc}
(D_+)_{mn}=e^{|k|l}\delta_{mn}  & , & (D_-)_{mn}=e^{-|k|l}\delta_{mn}  \\
D_+^{-1}=D_- & , & D_-^{-1}=D_+ \; .
\end{array}
\label{dpm}
\end{equation}

  Using Eqns. (\ref{nsmd},\ref{vta},\ref{mns},\ref{dpm}) one obtains:
  \begin{equation}
  \begin{array}{ccc}
  W^{-1} = (R_+V+P_+)^{-1}D_- & , & 
   X^{-1} = (R_+ +P_+T)^{-1}D_-
   \end{array}
   \end{equation}
  
Finally, Eqns. (\ref{eqa}) may be combined into the following eigenvalue problem for anti-symmetric modes:
\begin{equation}
0=
\left(
\begin{array}{cc}
 [M_dW^{-1} + h I][N_sX^{-1}+h I] & -\Omega^2 I  
\end{array}
\right) 
\left(
\begin{array}{c}
X H_e^a
\end{array}
\right)
\label{eqaf}
\end{equation}

Now, about symmetric modes: subtracting Eqns. (\ref{e1},\ref{e2}) and Eqns. (\ref{e3},\ref{e4}) between themselves and from Eqns. (\ref{ue}-\ref{do}), one obtains the following equations for this type of modes:

\begin{eqnarray}
0 & = &  (\Omega+h)[(M_+ +M_-)B_e^s
 -i(N_+ -N_-)B_o^a]
 \nonumber \\ &+ & (h+\Omega+1)[(N_+ -N_-)H_e^s
 -i(M_+ +M_-)H_o^a]
 \nonumber \\ \label{s1}  \\
 0 & = & -(\Omega-h) [(M_+ +M_-)B_e^s
 +i(N_+ -N_-)B_o^a) ]
 \nonumber \\ & +& (h-\Omega+1)[(N_+ -N_-)H_e^s
 +i(M_+ +M_-)H_o^a ] \nonumber \\
 \label{s2}  \\
 0 & = & P_- B_e^s -R_- H_e^s \label{s3} \\
  0 & = & R_- B_o^a -P_- H_o^a \label{s4}  
\end{eqnarray}

Summing and subtracting Eqns. (\ref{s1},\ref{s2}) one obtains:
 \begin{eqnarray}
 0 & = & -\Omega (N_dB_o^a+M_sH_o^a)
 -ih(M_sB_e^s+N_dH_e^s)-iN_dH_e^s
 \nonumber \\ \label{sf1} \\
 0 &= &   -\Omega (M_sB_e^s+N_dH_e^s)
 +ih (N_dB_o^a+M_sH_o^a) +iM_s H_o^a
  \nonumber \\ \label{sf2}
  \end{eqnarray}
  where 
   \begin{equation}
  \begin{array}{ccc}
   N_d \equiv N_+ -N_-$, $M_s \equiv M_+ +M_-
   \label{ndms}
   \end{array}
   \end{equation}
   
  Defining new variables:
  \begin{equation}
  \begin{array}{ccc}
  N_dB_o^a+M_sH_o^a=YH_o^a
  & , & M_sB_e^s+N_dH_e^s=ZH_e^s
  \label{nvs}
  \end{array}
  \end{equation}
  with, from Eqns. (\ref{vta},\ref{s3},\ref{s4},\ref{nvs}):
  \begin{equation}
  \begin{array}{ccc}
  Y \equiv N_dV+M_s
  & , & Z \equiv M_sT+N_d \; .
  \label{yzs}
  \end{array}
  \end{equation}
Now, from Eqns. (\ref{nsmd},\ref{vta},\ref{mns},\ref{ndms},\ref{yzs}) it results that $Y=W$, $Z=X$.
  The equations (\ref{sf1},\ref{sf2}) may then be written as:
  
  \begin{equation}
0=
\left(
\begin{array}{cc}
-\Omega I & -ih I-iN_dX^{-1} \\
iM_sW^{-1} + i h I& -\Omega I
\end{array}
\right) 
\left(
\begin{array}{c}
W H_o^a \\
X H_e^s
\end{array}
\right)
\label{eqs}
\end{equation}
These equations may be combined into the following eigenvalue problem for the symmetric modes:
\begin{equation}
0=
\left(
\begin{array}{cc}
 [M_sW^{-1} + h I][N_dX^{-1}+h I] & -\Omega^2 I  
\end{array}
\right) 
\left(
\begin{array}{c}
X H_e^s
\end{array}
\right)
\label{eqsf}
\end{equation}

\section{Scattering, reflection and transmission coefficients:}

Far from the region of geometric defects, the spin wave mode solutions correspond to those of flat surfaces (see section \ref{flat}). There 
a plane wave with a specified wavevector $k$,  and with related amplitudes of the magnetization and magnetostatic fields that vary along the thickness direction $y$, is a valid spin wave solution at a frequency $\omega (k)$ given by the dispersion relation of Eq. (\ref{DEf}). Thus, a scattering type of solution, with an incident, reflected and transmitted parts that are plane waves of wavevector $k$ may be regarded as a spin wave mode solution at frequency $\omega$, and far from the geometric defect its magnetostatic potential takes the following form:
\begin{equation}
\begin{array}{ccc}
\phi_{-\infty}=A_I f(y) e^{i(kx-\omega t)}+A_R g(y) e^{-i(kx+\omega t)}
& : & x \rightarrow -\infty \\
\phi_{\infty}=A_T f(y) e^{i(kx-\omega t)}
& : & x \rightarrow \infty \; .
\end{array}
\label{sso}
\end{equation}
Notice that the magnetostatic potential profiles of the incident and transmitted waves, $f(y)$, are the same, while that of the reflected wave, $g(y)$, satisfies $g(y)=f(-y)$, i.e. there is a mirror reflection symmetry with respect to the plane $y=0$ in the magnetostatic potential between right and left propagating spin wave modes (shape non reciprocity, see section \ref{flat}). 

Now we discuss the reflection and transmission of energy due to scattering, which allows to evaluate reflection (R) and transmission coefficients (T). Due to conservation of energy we will verify that $R+T=1$. According to Refs. \onlinecite{Akhiezer1968,Gupta1979} in the magnetostatic approximation the following expression represents the averaged over time energy current density $<\vec{F}>$ of spin waves (it is basically the electromagnetic Poynting vector; a time dependence of the fields as $\exp(-i\omega t)$ has been assumed):
\begin{equation}
 <\vec{F}>= -\frac{c \omega}{8\pi} 
 Re(i \phi^* \vec{b}) \; ,
 \label{Fav}
 \end{equation}
with $\phi$ and $\vec{b}=\vec{h}_D+4\pi \vec{m}$ the magnetostatic potential and magnetic induction associated with the spin wave, respectively. 
Indeed local conservation of energy is represented by a continuity equation of the form:
\begin{equation}
0= \partial u/\partial t +\nabla \cdot \vec{F} \; , 
\label{eco}
\end{equation}
with $u$ the energy density. We integrate the previous continuity equation over the volume $V$ of the film for a stationary process occurring at frequency $\omega$, and we average over time, the integration over $\partial u/\partial t$ averaged over time is zero (stationary process), and we get through Gauss's theorem:
\begin{equation}
0=\int_V \nabla \cdot <\vec{F}> =
\int_S <\vec{F}> \cdot d\vec{S}
\label{inco}
\end{equation}
The integration over the upper and lower surfaces of the film is zero (given the form of $<\vec{F}>$ from Eq. (\ref{Fav}) and a similar analysis to what we presented for the Extinction theorem in section \ref{exi}). Thus, from the remaining integration over the left and right cross sections of the film, Eq. (\ref{inco}) implies:
\begin{equation}
\int dz \int dy <F_x>_{-\infty}=\int dz \int dy <F_x>_{\infty} \; ,
\label{Fx}
\end{equation}
where the integration is over the cross sections at $x \rightarrow \mp \infty$ of the film (due to translation invariance in the $z$ direction, the integration in that direction cancels out). Thus, in Eq. (\ref{Fx}) the equal sign has come from conservation of energy (Eq. (\ref{eco})). The integrals in Eq. (\ref{Fx}) may be done using Eqns. (\ref{bxh},\ref{Fav}). Indeed:
\begin{equation}
\int dy <F_x>=-\frac{c\omega}{16\pi}
\int dy [ 2 \mu k |\phi|^2+\nu 
\frac{\partial}{\partial y}(|\phi|^2) ]
\label{Fxy}
\end{equation}
Then, according to Eqns. (\ref{sso},\ref{Fxy}):
\begin{eqnarray}
& & \int dy <F_x>_{\infty}=
\nonumber \\ & &
-|A_T|^2\frac{c\omega}{16\pi}
\{ 2 \mu k w [|f|^2]+\nu 
(|f(l)|^2-|f(-l)|^2) \} \; ,
\label{Fxp}
\end{eqnarray}
with:
\begin{equation}
w [|f|^2] \equiv \int_{-l}^l dy |f(y)|^2
\end{equation} 
Furthermore, 
\begin{eqnarray}
&  & \int dy <F_x>_{-\infty} = \nonumber \\
& & 
 -|A_I|^2\frac{c\omega}{16\pi}
\{ 2 \mu k w [|f|^2]+\nu 
(|f(l)|^2-|f(-l)|^2) \}
 \nonumber \\
& & 
-|A_R|^2\frac{c\omega}{16\pi}
\{ -2 \mu k w [|g|^2]+\nu 
(|g(l)|^2-|g(-l)|^2) \}
 \; .
\label{Fxm}
\end{eqnarray}
Using Eqns. (\ref{Fx},\ref{Fxp},\ref{Fxm}), and that $[|g|^2]=[|f|^2]$, $g(\pm l)=f(\mp l)$, one gets:
\begin{equation}
|A_I|^2-|A_R|^2=|A_T|^2
\end{equation}
or defining $|A_R/A_I|^2=R$, and $|A_T/A_I|^2=T$ as reflection and transmission coefficients, one gets:
\begin{equation}
R+T=1
\end{equation}
as expected due to energy conservation.

\subsection{Even and odd modes in 1D, phase shifts: \label{eou}}

A convenient way to handle numerically the previous scattering problem is to introduce mode solutions at frequency $\omega$ with parity properties. 

First, we present a simple scattering problem for a scalar field in 1D, $\psi (x)$, as a useful introduction. The incident, reflected and transmitted solutions far from the scatterer in this case are:

\begin{equation}
\begin{array}{ccc}
\psi_{-\infty}=A_I  e^{i(kx-\omega t)}+A_R  e^{-i(kx+\omega t)}
& : & x \rightarrow -\infty \\
\psi_{\infty}=A_T e^{i(kx-\omega t)}
& : & x \rightarrow \infty \; ,
\end{array}
\label{sss}
\end{equation}
or equivalently:
\begin{eqnarray}
\psi_{-\infty} & = & e^{-i\omega t} \{ (A_I  +A_R)\cos (kx)+ (A_I  -A_R)i\sin (kx)  \}
 \nonumber \\
\psi_{\infty} & = & 
e^{-i\omega t} 
A_T \{ \cos (kx)+i \sin (kx) \}
 \; .
 \label{ssx}
\end{eqnarray}

In terms of modes with symmetry properties, the scattering solutions of Eq. (\ref{sss}), that are associated with a wavevector $k$, read:

\begin{eqnarray}
\psi_{-\infty} & = & e^{-i\omega t} \{C_e \cos (kx-\delta_e) + C_o \sin (kx-\delta_o) \}
\nonumber \\
\psi_{\infty} & = & e^{-i\omega t} \{ C_e \cos (kx+\delta_e) + C_o \sin (kx+\delta_o) \}
 \; ,
 \label{seo}
\end{eqnarray}
where $\delta_e$ and $\delta_o$ are phase shifts produced by the scatterer in the even and odd solutions.
From Eqns. (\ref{ssx}) and (\ref{seo}) by simple algebra one gets:
\begin{equation}
\begin{array}{ccc}
A_I = -iC_oe^{-i\delta_o} & , &  A_R =  C_oe^{i\delta_e}\sin (\delta_e-\delta_o) 
 \\  A_T =  -iC_oe^{i\delta_e}\cos (\delta_e-\delta_o) & , & 
 C_e=-iC_o e^{i(\delta_e-\delta_o)}
\; ,
\label{ascs}
\end{array}
\end{equation}
from which it follows that:
\begin{equation}
\begin{array}{ccc}
 |A_R/A_I|^2=\sin^2(\delta_e-\delta_o) 
& , & |A_T/A_I|^2=\cos^2(\delta_e-\delta_o)
\end{array}
\end{equation}
Thus, the reflection and transmission coefficients may be determined through the difference of phase shifts associated with even and odd modes. 

\subsection{Film modes with symmetry properties, phase shifts:}

 The scattering modes solutions for the magnetostatic potential $\phi_{-\infty}(x,y,t)$ of Eqns. (\ref{sso}) at $x \rightarrow -\infty$ may be written as:
 
\begin{eqnarray}
\phi_{-\infty} & = & e^{-i \omega t} \{ A_I f(y) e^{ikx}+A_R f(-y) e^{-ikx} \}
\nonumber \\
& = & 
e^{-i \omega t} \{
e(y) [(A_I  +A_R)\cos (kx)+ (A_I  -A_R)i\sin (kx)]  \nonumber \\
& & +o(y) [(A_I  -A_R)\cos (kx)+ (A_I  +A_R)i\sin (kx)] \} \; ,
\nonumber \\
\label{fms}
\end{eqnarray}
with 
\begin{equation}
\begin{array}{ccc}
e(y) \equiv (f(y)+f(-y))/2 & , & o(y) \equiv (f(y)-f(-y))/2
\end{array}
\end{equation}
even and odd functions along the thickness of the film respectively. Similarly:

\begin{eqnarray}
\phi_{\infty} & = & e^{-i \omega t} A_T f(y) e^{ikx}
\nonumber \\
& = & 
e^{-i \omega t} A_T 
[e(y)+o(y)] [\cos (kx)+ i\sin (kx)]  \; .
\nonumber \\
\label{fps}
\end{eqnarray}

A convenient way to handle numerically the previous scattering problem is to introduce mode solutions at frequency $\omega$ with symmetry properties, similarly as in Eqns. (\ref{seo}):

\begin{eqnarray}
\phi_{-\infty} & = & e^{-i \omega t} \{
e(y) [C_e^e \cos (kx-\delta_e^e) + C_o^e \sin (kx-\delta_o^e)]  \nonumber \\
& & +o(y) [C_e^o \cos (kx-\delta_e^o) + C_o^o \sin (kx-\delta_o^o)] \} \; ,
\nonumber \\
\label{fmp}
\end{eqnarray}
\begin{eqnarray}
\phi_{\infty} & = &e^{-i \omega t} \{
e(y) [C_e^e \cos (kx+\delta_e^e) + C_o^e \sin (kx+\delta_o^e)]  \nonumber \\
& & +o(y) [C_e^o \cos (kx+\delta_e^o) + C_o^o \sin (kx+\delta_o^o] \} \; ,
\nonumber \\
\label{fpp}
\end{eqnarray}
i.e. we are distinguishing even-even ($C_e^e$), even-odd ($C_e^o$), odd-even ($C_o^e$) and odd-odd ($C_o^o$) modes, considering parity properties in the $y$ and $x$ directions respectively. Similarly to what was done in the previous section, comparing Eqns. (\ref{fms},\ref{fmp}) and Eqns. (\ref{fps},\ref{fpp}) for the different types of modes, one obtains:
\begin{equation}
\begin{array}{ccc}
A_I =-iC_o^e e^{-i\delta_o^e} & , &  A_R =  C_o^e e^{i\delta_e^e}\sin (\delta_e^e-\delta_o^e)
  \\  A_T =  -iC_o^e e^{i\delta_e^e}\cos (\delta_e^e-\delta_o^e) & , & 
 C_e^e=-iC_o^e e^{i(\delta_e^e-\delta_o^e)}
\; ,
\label{ase}
\end{array}
\end{equation}

and 
\begin{equation}
\begin{array}{ccc}
A_I =-iC_o^o e^{-i\delta_o^o} & , &  A_R =  -C_o^o e^{i\delta_e^o}\sin (\delta_e^o-\delta_o^o)
 \\  A_T =  -iC_o^o e^{i\delta_e^o}\cos (\delta_e^o-\delta_o^o) & , & 
 C_e^o=-iC_o^o e^{i(\delta_e^o-\delta_o^o)}
\; ,
\label{aset}
\end{array}
\end{equation}

Notice that Eqns. (\ref{ase},\ref{aset}) are consistent if $C_o^o=C_o^ee^{i(\delta_e^e-\delta_o^e)}$, $\delta_e^o=\delta_o^e$ and $\delta_o^o=\delta_e^e$. Indeed calling $\delta_e^e=\delta_s$, $\delta_o^e=\delta_a$, $C_o^e=C_a$, then: 
\begin{eqnarray}
A_I & = & -iC_a e^{-i\delta_a} \nonumber \\ 
A_R & = &  C_a e^{i\delta_s}\sin (\delta_s-\delta_a)  \nonumber \\ 
A_T & = &  -iC_a e^{i\delta_s}\cos (\delta_s-\delta_a)
\; . 
\label{asef}
\end{eqnarray}
Also, notice that $C_ae^{i \delta_s}=iC_s e^{i \delta_a}$, with $C_s \equiv C_e^e$. 
Then Eqns. (\ref{fmp},\ref{fpp}) take the form:

\begin{eqnarray}
\phi_{-\infty} & = & A_Ie^{-i \omega t} \{
e(y) [e^{i\delta_s} \cos (kx-\delta_s) + i e^{i\delta_a}\sin (kx-\delta_a)]  \nonumber \\
& & +o(y) [e^{i\delta_a} \cos (kx-\delta_a) +i e^{i\delta_s }\sin (kx-\delta_s)] \} \; ,
\nonumber \\
\label{fmpu}
\end{eqnarray}
\begin{eqnarray}
\phi_{\infty} & = & A_Ie^{-i \omega t} \{
e(y) [e^{i\delta_s} \cos (kx+\delta_s) + i e^{i\delta_a} \sin (kx+\delta_a)]  \nonumber \\
& & +o(y) [e^{i\delta_a} \cos (kx+\delta_a) +  ie^{i\delta_s} \sin (kx+\delta_s)] \} \; ,
\nonumber \\
\label{fppu}
\end{eqnarray}
Then, the previous solutions at $x = \pm \infty$ may be written as:
\begin{eqnarray}
\phi_{-\infty} & = & A_Ie^{-i \omega t} \{e^{i\delta_s}
[e(y) \cos (kx-\delta_s) + i o(y) \sin (kx-\delta_s)]  \nonumber \\
& & +e^{i\delta_a} [o(y) \cos (kx-\delta_a) +i e(y) \sin (kx-\delta_a)] \} \; ,
\nonumber \\
\label{fmpf}
\end{eqnarray}
\begin{eqnarray}
\phi_{\infty} & = &  A_Ie^{-i \omega t} \{e^{i\delta_s}
[e(y) \cos (kx+\delta_s) + i o(y) \sin (kx+\delta_s)]  \nonumber \\
& & +e^{i\delta_a} [o(y) \cos (kx+\delta_a) +i e(y) \sin (kx+\delta_a)] \}
\nonumber \\
\label{fppf}
\end{eqnarray}
Thus, the solutions involve sums of symmetric and anti-symmetric solutions under the inversion symmetry of the system, i.e. $(x,y) \rightarrow -(x,y)$. These symmetric and anti-symmetric solutions have clearly phase shifts $\delta_s$ and $\delta_a$ at $|x| \rightarrow \infty$, respectively. 
Furthermore, from Eqns. (\ref{asef}) one concludes that:
\begin{eqnarray}
 R & = & |A_R/A_I|^2 = \sin^2(\delta_s-\delta_a)  \nonumber \\
T & = & |A_T/A_I|^2 = \cos^2(\delta_s-\delta_a) \; .  \nonumber \\
\label{RT}
\end{eqnarray}
Thus, the reflection (R) and transmission (T) coefficients may be calculated using the differences of the phase shifts of the just mentioned symmetric and anti-symmetric modes, phase shifts that are produced by the geometric defects.

Notice that for a film with flat surfaces these symmetric (S) and anti-symmetric (A) mode solutions for the magnetostatic potential are:
\begin{eqnarray}
S(x,y) & = & 
e(y) \cos (kx) + i o(y) \sin (kx)  \nonumber \\
A(x,y) & = & o(y) \cos (kx) +i e(y) \sin (kx)
\label{SA}
\end{eqnarray}
These are simply interpreted, since they are basically the sum and difference of right and left traveling spin wave modes (here we write just the magnetostatic potential of the spin wave):
\begin{eqnarray}
S(x,y) & = & (f(y) e^{ikx}+f(-y) e^{-ikx})/2  \nonumber \\
A(x,y) & = & (f(y) e^{ikx}-f(-y) e^{-ikx})/2
\label{SAi}
\end{eqnarray}

\section{Examples of magnetostatic surface scattering:}

\begin{figure}[ht]
     \centering
    \includegraphics[width=.4\textwidth]{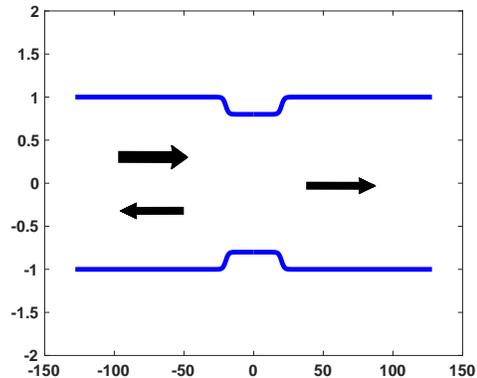}
    \caption{Geometric defects chosen: depressions symmetrically located. Approximate dimensions of the depressions in this figure: depth $p=0.1$, length $2a=40$.}
    \label{fig:GeomDep}
\end{figure}

In section \ref{ssm} the geometric defects were presented as modifying the upper and lower surfaces of the film as $y=l+\eta (x)$, and $y=-l+\xi (x)$, respectively. In order to present examples of magnetostatic surface scattering using the theory presented, we chose a symmetric configuration with $\eta (x) = -\xi (x)= p [-1+\tanh((x-a)/b)]$ (formulae valid for $x \geq 0$, and with reflection symmetry with respect to $x=0$). Fig. \ref{fig:GeomDep} represents the film with these depressions for the case $a=20$ (we also did calculations with $a=10$), and we took $b=2$, $p=0.1$. These correspond to symmetrically located depressions.

The calculations that will be presented in the following are done for films of length between $-L \leq x \leq L$, with $L=1024\Delta$, so Fig. \ref{fig:GeomDep} is just a section of the full length, shown for illustrative purposes. $\Delta$ is a unit of length, so wave-vectors will have units of $1/\Delta$. The thickness of the film is taken between $y=-\Delta$ and $y=\Delta$. All the results that we present have no units, for example for the coordinate $x$ actually we plot $x/\Delta$: indeed magnetostatic equations are independent of $\Delta$, there is no underlying length scale. Also, we take $h_0=H_0/4\pi M_s=0.2$, so that the range of frequencies of the Damon-Eshbach modes in this case is between $\Omega_l=\sqrt{h_0(h_0+1)} \simeq 0.49$ and $\Omega_u= h_0+1/2=0.7$.  

First, we discuss results on scattering of propagating magnetostatic surface modes. As presented in the theory sections, our theory is able to calculate the scattering of an incident magnetostatic surface wave of wave-vector $k$, with an associated wave-length $\lambda = 2\pi/k$. Reflection ($R$)  and transmission coefficients ($T$) of this incident surface wave were determined through phase shifts as shown in Eqns. (\ref{RT}). Then, the transmission coefficient is given by $T= \cos^2 \delta$, where we have defined $\delta \equiv \delta_s-\delta_a$. In the following Fig. \ref{fig:T} the transmission coefficient $T$ is shown as a function of frequency (frequency $\Omega \equiv \omega/4\pi M_s |\gamma|$ and wave-vector $k$ are related through Eq. (\ref{DEf}) for these scattering modes) for the two examples of depressions. 

\begin{figure}[ht]
     \centering
    \includegraphics[width=.4\textwidth]{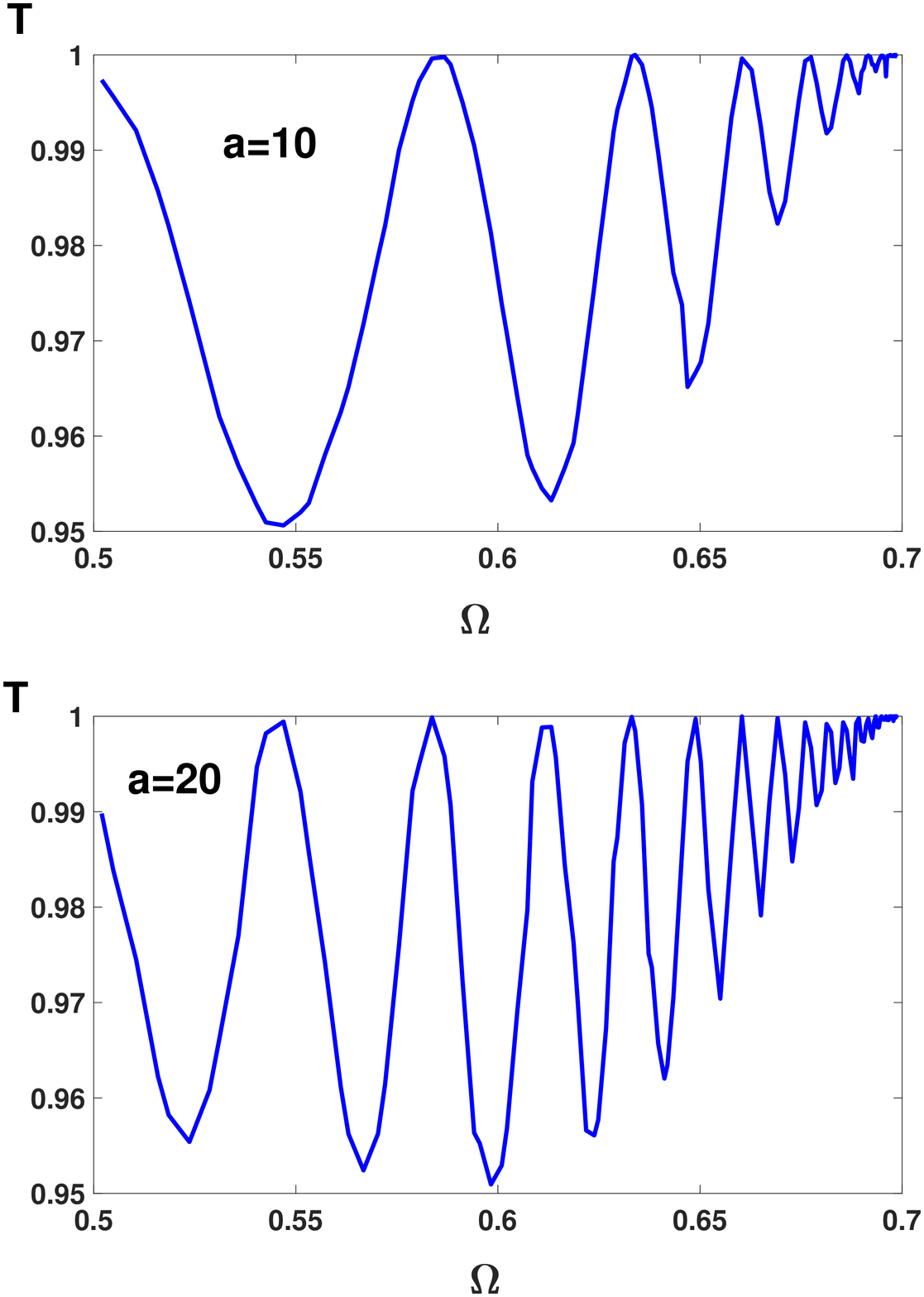}
    \caption{Scattering transmission coefficient $T$ as a function of surface modes frequency $\Omega$, for depression half-widths $a=10$ and $a=20$ respectively ($h_0=0.2$).}
    \label{fig:T}
\end{figure}
Thus, these particular obstacles do not affect much the overall transmission ($T$ hovers between 0.95 and 1), but interestingly there are particular frequencies at which there is perfect transmission ($T=1$), which we associate with resonances \cite{Sprung1996}. Resonances have been qualitatively associated with a wave being trapped for a while through successive reflections in a “potential well", and then escaping from it. As can be seen in Fig. \ref{fig:T}, as the length of the depression is larger more “resonances" occur, which is reasonable given the previous interpretation.

The previous transmission coefficients were calculated by determining the phase shifts $\delta_s$, $\delta_a$, associated with the symmetric and anti-symmetric modes under inversion. In the following we comment about the determination of these phase shifts with our method. The phase shifts $\delta_a$, $\delta_s$ are determined by plotting the inverse cosine transforms associated with the even functions $H_e^a$ and $H_e^s$ of Eqns. (\ref{eqaf}) and (\ref{eqsf}), respectively. In both cases (symmetric and anti-symmetric modes) we plot the inverse Fourier cosine transforms for no geometric defects and with them, and then we determine the phase shifts between these plots at a distance $x \sim 100$ from the center $x=0$, i.e. approximately at the right end of Fig. \ref{fig:GeomDep}. Notice that the parameter $L$, that reflects the end of an actual film, and that needs to be taken as finite in a numerical calculation, is a “regulator" for the problem, i.e. it discretizes the number of modes, and allows to count them. Also, due to this “regulator" the inverse Fourier cosine transforms have zero derivatives at $x=L$ and the plots of $H_e^a(x) $ and $H_e^s(x)$ coincide at $x=L$ for the film with and without defects, meaning that the phase shifts determined at $x=L$ are zero, that's why in our case we determine these phase shifts closer to the defects, i.e. at approximately $x=L/10$. 

The following Fig. \ref{fig:Delta} plots the difference of phase shifts of symmetric and anti-symmetric modes, i.e. $\delta \equiv \delta_s-\delta_a$, as a function of the wavelength of the incident magnetostatic surface wave.
\begin{figure}[ht]
     \centering
    \includegraphics[width=.4\textwidth]{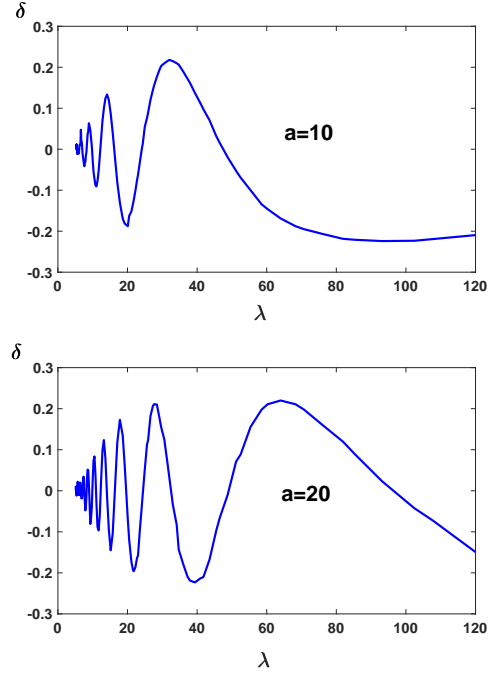}
    \caption{Scattering phase shift $\delta \equiv \delta_s-\delta_a$ as a function of surface modes wavelength $\lambda=2\pi/k$, for depression half-widths $a=10$ and $a=20$ respectively.}
    \label{fig:Delta}
\end{figure}

Clearly $\delta$ oscillates in a non-uniform way as a function of $\lambda$. Peaks of these oscillations represent  lower transmissions, and zeros represent full transmission or resonances. These oscillations of $\delta$ between positive and negative values of the curves of Fig. \ref{fig:Delta} imply directly the recurrence of full transmissions of Fig. \ref{fig:T}.

Now we turn to discuss the appearance of localized modes associated with the presence of these geometric defects which are depressions. Indeed for both extensions of the depressions, i.e. $a=10, 20$, and for the given depth, $p=0.1$, interestingly there are two localized modes appearing close to the region of the depressions. In both cases ($a=10, 20$) there is a mode of lower frequency that is localized mainly in the interior of the depressions, and a higher frequency mode localized in the interior and in the contiguous region to the depression. The following Fig. \ref{fig:FirstOddModes} is a plot of the shapes of $H_e^a$ for the lowest frequency localized anti-symmetric modes, with $a=10, 20$.

\begin{figure}[ht]
     \centering
    \includegraphics[width=.4\textwidth]{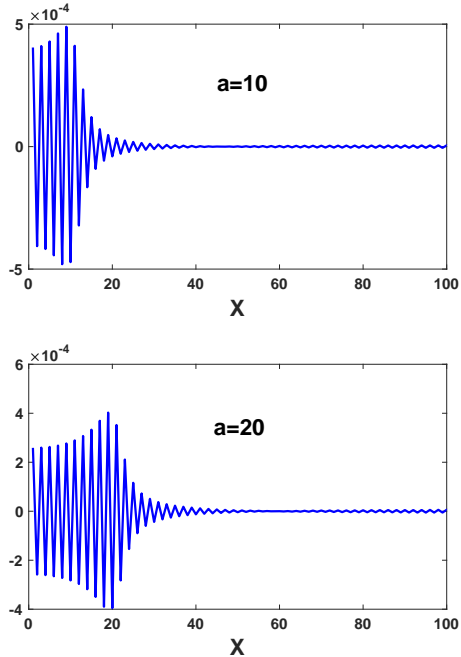}
    \caption{Shapes of the lowest frequency localized antisymmetric modes (variable $H_e^a$) as a function of the distance from the center $x=0$ of the film, for depression half-widths $a=10$ and $a=20$ respectively.}
    \label{fig:FirstOddModes}
\end{figure}

Clearly these previous modes are approximately localized to the regions limited by $x=10, 20$, when $a=10, 20$,  respectively. The following Fig. \ref{fig:SecondEvenModes} is a plot of the shapes of $H_e^s$ for the second lowest frequency localized symmetric modes, with $a=10, 20$:

\begin{figure}[ht]
     \centering
    \includegraphics[width=.4\textwidth]{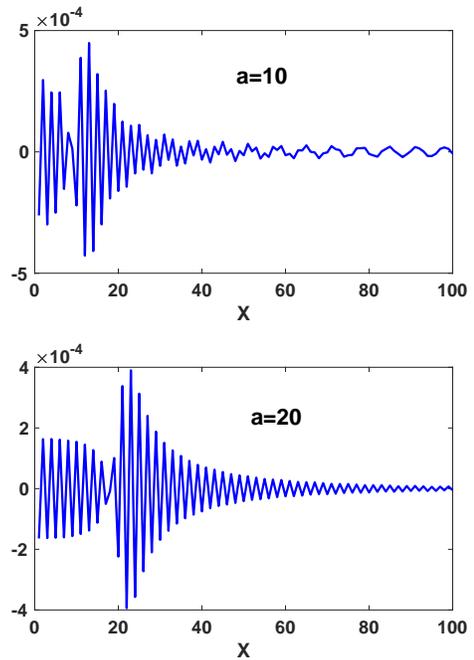}
    \caption{Shapes of the second lowest frequency localized symmetric modes (variable $H_e^s$) as a function of the distance from the center $x=0$ of the film, for depression half-widths $a=10$ and $a=20$ respectively.}
    \label{fig:SecondEvenModes}
\end{figure}
The previous “second" localized symmetric modes do have amplitudes  in the interior regions of the depressions, and decaying amplitudes outside the depressions of an extent similar to the interior regions.

The localized modes at the depressions do have cosine Fourier transform coefficients of higher amplitudes at the higher end of the wave-vector range, as evidenced in the following Fig. \ref{fig:FirstWavEvenModes}.

\begin{figure}[ht]
     \centering
    \includegraphics[width=.4\textwidth]{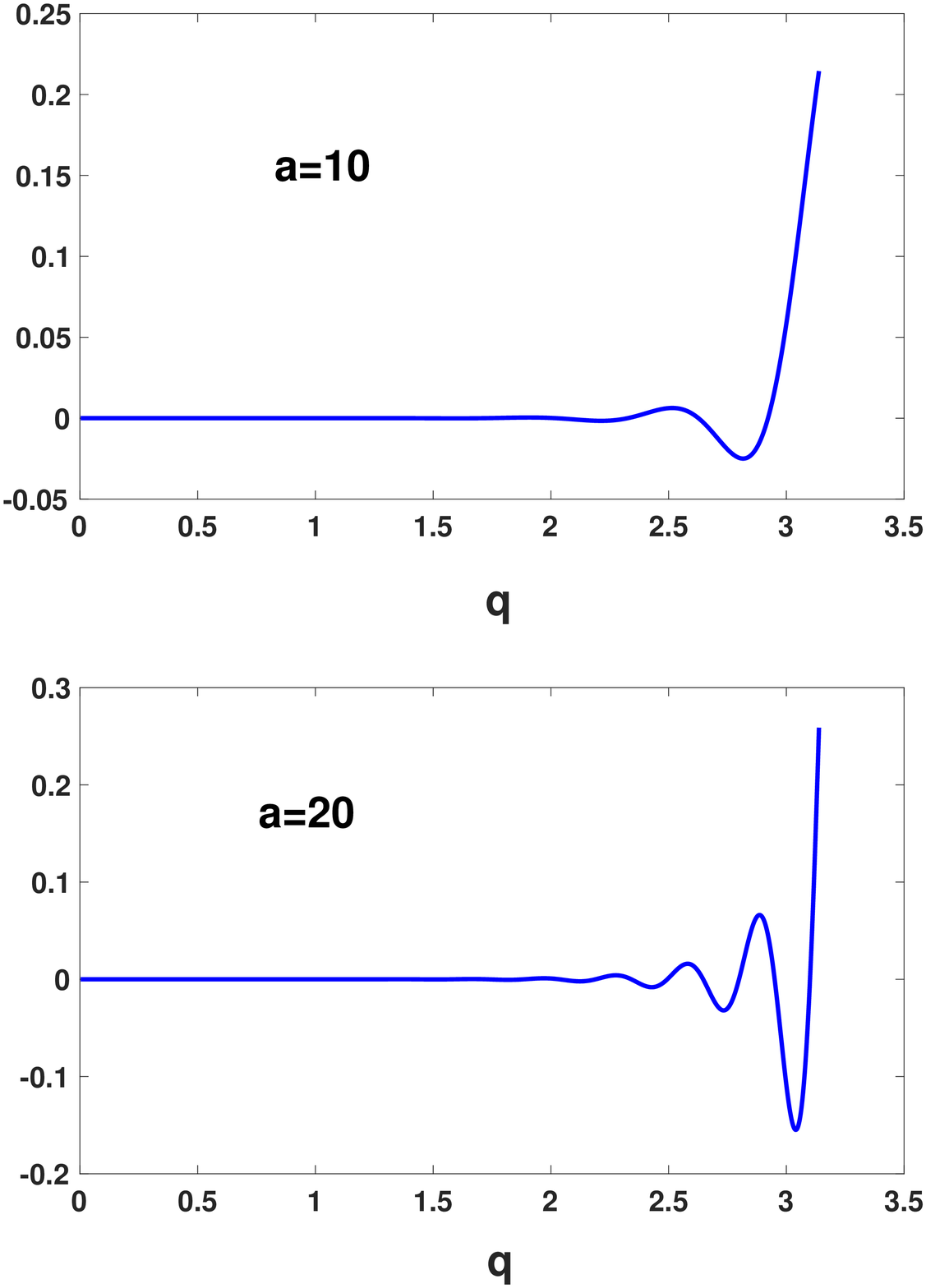}
    \caption{Cosine Fourier transform coefficients of the lowest frequency localized symmetric modes (variable $H_e^s$) as a function of the wavevector $q$, for depression half-widths $a=10$ and $a=20$ respectively.}
    \label{fig:FirstWavEvenModes}
\end{figure}

Thus, these modes do have a short wavelength content in their structure, which allows them to be localized at the depressions (they also oscillate quite a bit inside them). 

We did similar scattering calculations for a semi-infinite medium (theory explained in section \ref{smi}), and we did find that there are very similar localized modes at their surface when there is a single analogous depression. One would expect a finite thickness effect of the film if the depressions had a higher depth. All this is consistent with the short wavelength content of these localized modes. 

Furthermore in the following Fig. \ref{fig:FreqEven} the frequencies of the two symmetric localized modes are plotted in terms of a variable depth $p$ of the depressions.

\begin{figure}[ht]
     \centering
    \includegraphics[width=.4\textwidth]{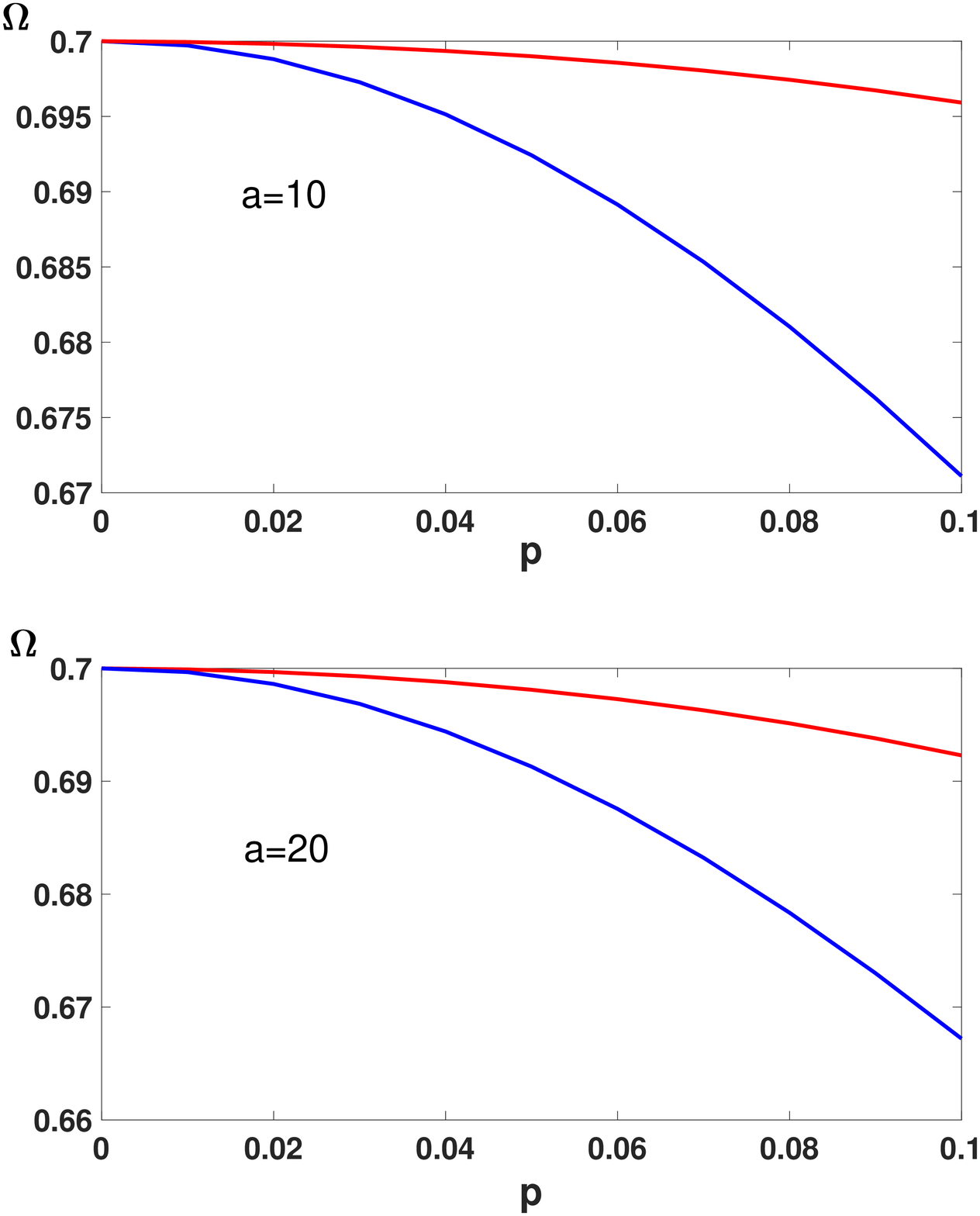}
    \caption{Frequencies, $\Omega$, of the localized symmetric modes as a function of the depth $p$ of the geometric depressions (the thickness of the film is $w=2$), for depression half-widths $a=10$ and $a=20$ respectively.}
    \label{fig:FreqEven}
\end{figure}

Thus, there is a monotonic decrease of the frequencies of these localized modes as the depth of the depressions increases. The decrease starts at the frequency of the surface modes of a perfect semi-infinite surface (or short wavelength limit of a perfect film also), i.e. $\Omega_u=h_0+1/2=0.7$. The anti-symmetric localized modes have similar frequencies of those of the symmetric modes shown in Fig. \ref{fig:FreqEven}. Notice that these discrete frequencies of localized modes are immersed in the continuum spectrum of the DE surface modes, as in other studies \cite{Herrick1975}.

We make a final comment about the calculations. The eigenvalue problems that are solved, i.e. Eqns. (\ref{eqaf}), (\ref{eqsf}), are diagonal when the film has no defects: then we start the calculations with $p=0$ or flat surfaces, the eigenvalues and eigenvectors are very simple, each one corresponds to a given value of $k$. Then the parameter $p$ is changed step by step (this allowed to plot for example the frequencies of Fig. (\ref{fig:FreqEven})), and the different eigenvalues and eigenvectors are followed as they evolve with $p$ growing: this allows to identify the effect of the defects in the scattering of a mode with associated wavevector $k$ at infinity. 

\section{Conclusions:}

A study of scattering of magnetostatic Damon-Eshbach (DE) waves by geometric defects was presented. A theory was developed that may be applied to defects of arbitrary shapes in ferromagnetic films and semi-infinite media. This theory is based on the Green-Extinction theorem: integral equations are obtained for the modes on the surfaces and their frequencies. We chose symmetrically located defects: this allowed to simplify the scattering calculations, that may be framed in terms of scattering phase shifts of symmetric and anti-symmetric modes under inversion (the phase shifts measure how much the solutions are displaced in the flat regions due to the defects with respect to the case without defects). We calculated transmission and reflection coefficients associated with energy conservation in the scattering process, these are directly written in terms of the mentioned phase shifts. We provided examples of the application of the theory by choosing as geometrical defects symmetrically located depressions: we varied a bit their extent and depth. The results are that the transmission coefficients decrease with depth of the depressions, and they show “resonances", i.e. there are particular incident wavelengths (or frequencies) at which there is perfect transmission, this is associated with a wave that is temporarily trapped in the depressions and then leaves. Interestingly we also found the appearance of two localized modes for both types of modes (symmetric and anti-symmetric under inversion)  in the depressions, with frequencies that are lower than the short wave-length limit of DE surface waves in films, i.e. they are bound states in the continuum. The lowest frequency modes are localized inside the depressions, while the higher frequency modes do have localization inside the depression, and to a similar extent also they do show amplitude outside the depressions. Very similar localized modes do appear in a semi-infinite medium with depressions, since these modes have a high content of short wavelengths in themselves. The latter indicates that these types of scattering effects should appear in all surfaces with roughness or more pronounced geometric defects.

\section{Acknowledgments}
 We acknowledge support by Fondecyt Project 1200829 (Chile), and ANID/PIA Basal Program for Centers of Excellence, Grant no. AFB 220001 CEDENNA (Chile). 
\label{sec:Ack}

\section{Appendix}
\label{sec:appendix}

\subsection{Continuous and discrete Sine and Cosine Fourier Transforms: \label{scFT}}

The continuous cosine and sine transforms of even, $E(x)$, and odd, $O(x)$, functions are defined as:
\begin{eqnarray}
E_c(q) & = & \int_{-\infty}^{\infty} dx \cos (qx)  E (x) 
 = 2\int_{0}^{\infty} dx \cos (qx)  E (x) 
 \label{cftd} \nonumber \\ \\
 O_s(q) & = & \int_{-\infty}^{\infty} dx \sin (qx)  O (x)= 2 \int_{0}^{\infty} dx \sin (qx) O(x) \; . 
\label{sftd} \nonumber \\
\end{eqnarray}
From these equations, it follows that $E_c(-q)=E_c(q)$, and $O_s(-q)=-O_s(q)$. These may be inverted as follows:
\begin{eqnarray}
E (x) & = & 
\frac{1}{2\pi} \int_{-\infty}^{\infty} dq \cos (qx) E_c (q) =
\frac{1}{\pi} \int_{0}^{\infty} dq \cos (qx) E_c (q) 
\label{icfta} \nonumber \\ \\
O (x) & = &
\frac{1}{2\pi} \int_{-\infty}^{\infty} dq \sin (qx) O_s (q) =
\frac{1}{\pi} \int_{0}^{\infty} dq \sin (qx) O_s (q) \; . 
\label{isfta} \nonumber \\
\end{eqnarray} 
A connection of these mentioned continuous sine and cosine Fourier transforms of Eqns. (\ref{cftd}-\ref{isfta}) with the usual discrete sine and cosine transforms is the following:
\begin{eqnarray}
\frac12 E_c(q_k) & = &  \int_0^{\infty} dx \cos (q_k x) E(x) =\Delta E_k \nonumber \\ & = & \Delta \{ \frac{1}{2}(E_0+E_N(-1)^k) +\sum_{j=1}^{N-1}  \cos (q_k j \Delta) E_j \} \nonumber \\ 
\label{dcft} \\
\frac12 O_s(q_k) & = &  \int_0^{\infty} dx \sin (q_k x) O(x) = \Delta O_k \nonumber \\ & = & \Delta \sum_{j=1}^{N-1}  \sin (q_k j \Delta) O_j  \; ,
\label{dsft}
\end{eqnarray}
where the points located between $0 < x < L$ are discretized as follows: $x_j = j \Delta$, $j=0, \ldots, N$, with $L=N\Delta$. Also, $q_k= \pi k/N \Delta=\pi k/L$.

Following the continuous formulae, the inverse discrete cosine and sine transforms are the following:

\begin{eqnarray}
 E(x_j) & = & E_j= \frac{1}{\pi} \int_0^{\infty} dq \cos (q x_j) E_c(q) \nonumber \\ & = & \frac{2}{N} \{ \frac{1}{2}(E_{k=0}+E_{k=N}(-1)^j) +\sum_{k=1}^{N-1}  \cos (q_k j \Delta) E_k \}  \nonumber \\
 \label{idcft} \\
O(x_j) & = & \frac{1}{\pi} \int_0^{\infty} dq \sin (q x_j) O_s(q) = \frac{2}{N} \sum_{k=1}^{N-1}  \sin (q_k j \Delta) O_k 
\nonumber \\
\label{idsft}
\end{eqnarray}
with $q_k j \Delta=\pi k j/N$.

  \subsection{Even obstacles, thin films: \label{sam}}
  
By writing the fields in terms of even and odd functions with respect to $x =0$, the Extinction Eqns. (\ref{ef}) lead to  ($B(x)=B_e(x)+B_o(x)$, $\Phi (x)=\Phi_e(x)+\Phi_o(x)$):
\begin{eqnarray}
0 & = &  \int_{-\infty}^{\infty} dx e^{-|k| \eta (x)}
\{ \cos (k x)B_e^u(x) 
\nonumber \\ & + & 
[k \sin  (kx) \eta'(x)-|k|\cos (kx)]\Phi^u_e(x) -i\sin (k x)B_o^u(x)
\nonumber \\ & + & 
i [k \cos  (kx) \eta'(x)+|k|\sin (kx)]\Phi^u_o(x)
 \} \label{uex} \\
0 & = &
\int_{-\infty}^{\infty} dx e^{|k| \xi (x)}
\{ \cos (k x)B_e^d(x)
\nonumber \\ & - & 
[k \sin  (kx) \xi'(x)+|k|\cos (kx)]\Phi^d_e(x) -i\sin (k x)B_o^d(x)
\nonumber \\ & - & 
i [k \cos  (kx) \xi'(x)-|k|\sin (kx)]\Phi^d_o(x)
\} \label{dex} \\
0 & = & 
\int_{-\infty}^{\infty} dx e^{\pm |k| (l+\eta (x))} \{ \cos (kx)
B_e^u(x)-i \sin (kx)B_o^u(x)
\nonumber \\ & + &
[\cos (kx)(\nu k \pm \mu |k|)+\sin (kx) \eta'(x)(\mu k \pm \nu |k|)]\Phi^u_e(x) 
\nonumber \\ & - &
i[\sin (kx)(\nu k \pm \mu |k|)-\cos (kx) \eta'(x)(\mu k \pm \nu |k|)]\Phi^u_o(x)
\} \nonumber \\ & + & 
\int_{-\infty}^{\infty} dx e^{\pm |k| (-l+\xi (x))} \{ \cos (kx)
B_e^d(x)-i\sin (kx)B_o^d(x)
\nonumber \\ & -&
[\cos (kx)(\nu k \pm \mu |k|)+ \sin (kx) \xi'(x)(\mu k \pm \nu |k|)]\Phi^d_e(x) 
\nonumber \\ &+ &
i[\sin (kx)(\nu k \pm \mu |k|)- \cos (kx) \xi'(x)(\mu k \pm \nu |k|)]\Phi^d_o(x)
\nonumber \\
\label{mex}
\end{eqnarray}

Using the representation of the unknown fields in terms of inverse Sine and Cosine Fourier Transforms (see next Appendix section \ref{scFT}) leads to:  
\begin{eqnarray}
0 & = &  B_e^u(k) -|k| \Phi_e^u(k)  -is(k)B_o^u(k) +ik \Phi_o^u(k)
\nonumber \\ & & +\frac{1}{\pi} \int_{0}^{\infty} dq  \{ C_u^{-|k|}(k,q) B_e^u(q) -q 
S_u^{-|k|}(k,q) \Phi_e^u(q) \}
\nonumber \\ & &
-\frac{i}{\pi} \int_{0}^{\infty} dq  \{ s(k) S_u^{-|k|}(k,q) B_o^u(q) 
-q s(k)C_u^{-|k|}(k,q) \Phi_o^u(q)  \}
 \label{ut}  \nonumber \\ \\
 0 & = & B_e^d(k) -|k| \Phi_e^d(k) -is(k)B_o^d(k) +ik\Phi_o^d(k)
 \nonumber \\ & &
 + \frac{1}{\pi} \int_{0}^{\infty} dq  \{  C_d^{|k|}(k,q) B_e^d(q) -q
S_d^{|k|}(k,q) \Phi_e^d(q) \}
 \nonumber \\ & &
 -\frac{i}{\pi} \int_{0}^{\infty} dq  \{ s(k) S_d^{|k|}(k,q) B_o^d(q) 
  -q s(k)C_d^{|k|}(k,q) \Phi_o^d(q) \}
 \label{dt}  \nonumber \\ \\
0 & = & e^{\pm |k|l}[
B_e^u(k)-is(k)B_o^u(k)
\nonumber \\ &  &
+(k\nu \pm |k|\mu) (\Phi_e^u(k)-is(k)\Phi_o^u(k))]
\nonumber \\ &  & +
\frac{e^{\pm |k|l} }{\pi}  \int_{0}^{\infty} dq  \{ C_u^{\pm|k|}(k,q) B_e^u(q)-is(k) S_u^{\pm|k|}(k,q) B_o^u(q)
\nonumber \\ & &
+q(\nu \pm s(k)\mu) [s(k) S_u^{\pm|k|}(k,q) \Phi_e^u(q) 
-iC_u^{\pm|k|}(k,q) \Phi_o^u(q)] \} 
\nonumber \\
&  & + e^{\mp |k|l}[
B_e^d(k)-is(k)B_o^d(k)
\nonumber \\ & & 
-(k\nu \pm |k|\mu) (\Phi_e^d(k)-i s(k)\Phi_o^d(k))]
\nonumber \\ & & +
\frac{e^{\mp|k|l}}{\pi}  \int_{0}^{\infty} dq  \{ C_d^{\pm|k|}(k,q) B_e^d(q)-i s(k) S_d^{\pm|k|}(k,q) B_o^d(q)
\nonumber \\ & &
-q(\nu \pm s(k)\mu) [s(k) S_d^{\pm|k|}(k,q) \Phi_e^d(q) 
-iC_d^{\pm|k|}(k,q) \Phi_o^d(q)] \} 
\; , 
\nonumber \\ 
\label{mt}
\end{eqnarray}

 Indeed, the following equations help to understand the previous Eqns. (\ref{ut}-\ref{mt}):

\begin{eqnarray}
    f_{\eta}^{\pm}(x) & \equiv & e^{\pm |k| \eta (x)}(k\sin (kx) \eta' \pm |k| \cos (kx)) 
    \nonumber \\ & & = 
    \pm s(k) \frac{\partial}{\partial x}[\sin (kx)e^{\pm |k| \eta (x)} ]
\end{eqnarray}

Then,

\begin{eqnarray}
    \int_{-\infty}^{\infty}dx \cos (q x) f_{\eta}^{\pm}(x)  & = & 
    \pm q s(k)[\pi(\delta(k-q)-\delta(k+q))
 \nonumber \\ & &    
    +S_u^{\pm |k|}(k,q)]
    \label{ife}
\end{eqnarray}

Similarly,
\begin{eqnarray}
 \int_{-\infty}^{\infty} \cos (q x) f_{\xi}^{\pm}(x)  & = &
    \pm q s(k)[\pi(\delta(k-q)-\delta(k+q))
    \nonumber \\ & & +S_d^{\pm |k|}(k,q)]
\end{eqnarray}

 Notice that
\begin{eqnarray}
 & [\cos (kx)(\nu k \pm \mu |k|)+\sin (kx) \eta'(x)(\mu k \pm \nu |k|)]  \nonumber \\ &
 =
 (\mu\pm s(k) \nu)(k\sin (kx)\eta'(x) \pm |k| \cos (kx)) \; ,
\end{eqnarray}
and it leads through Eq. (\ref{ife}) to:
\begin{eqnarray}
 & & \int_{-\infty}^{\infty} dx \cos (qx)
 e^{\pm |k|\eta (x)}[\cos (kx)(\nu k \pm \mu |k|)
 \nonumber \\ & & 
 +\sin (kx) \eta'(x)(\mu k \pm \nu |k|)] 
 \nonumber \\ & = &
 q (\nu \pm s(k) \mu)[\pi(\delta(k-q)-\delta(k+q))+S_u^{\pm |k|}(k,q)] 
  \nonumber \\
\end{eqnarray}

Also,
\begin{eqnarray}
    g_{\eta}^{\pm}(x) & \equiv & e^{\pm |k| \eta (x)}(k\cos (kx) \eta' \mp |k| \sin (kx)) 
    \nonumber \\
    & = &  
    \pm s(k) \frac{\partial}{\partial x}[\cos (kx)e^{\pm |k| \eta (x)} ] \;  .
\end{eqnarray}

Then,

\begin{eqnarray}
    \int_{-\infty}^{\infty} dx \sin (q x) g_{\eta}^{\pm}(x)  
    & = &  
    \mp q s(k)[\pi(\delta(k-q)+\delta(k+q))
    \nonumber \\ & & 
    +C_u^{\pm |k|}(k,q)] \; .
    \label{ige}
\end{eqnarray}
Also, 
\begin{eqnarray}
 & [\sin (kx)(\nu k \pm \mu |k|)-\cos (kx) \eta'(x)(\mu k \pm \nu |k|)] 
  \nonumber \\
 & = -
 (\mu\pm s(k) \nu)(k\cos (kx)\eta'(x) \mp |k| \sin (kx)) \; ,
\end{eqnarray}
and it leads through Eq. (\ref{ige}) to
\begin{eqnarray}
 & & \int_{-\infty}^{\infty} dx e^{\pm |k|\eta(x)} \sin (qx) [\sin (kx)(\nu k \pm \mu |k|)
 \nonumber \\
 & & 
 -\cos (kx) \eta'(x)(\mu k \pm \nu |k|)]
 \nonumber \\
 & = &
 q(\nu \pm \mu s(k))[\pi(\delta(k-q)+\delta(k+q))+C_u^{\pm |k|}(k,q)] \; .
  \nonumber \\
\end{eqnarray}

\subsection{Semi-infinite medium: \label{smi}}

We consider that the semi-infinite magnetized medium is located in $y \geq 0$, and that the geometric perturbation of the lower surface ($y \simeq 0$) is $\xi (x) = -\eta (x)$, i.e. $y(x) = -\eta (x)$ represents the perturbed surface. 
One may obtain Extinction equations for semi-infinite mediums as special cases of the Extinction equations (\ref{dtd},\ref{mtd}):

\begin{eqnarray}
 0 & = & B_e(k) -is(k)B_o(k) -H_e(k) 
 +is(k)H_o(k) 
 \nonumber \\ & +&
 \frac{2}{N} \sum_{q}  \{  C_-(k,q) [ B_e(q) 
 +i s(k) H_o(q) ] \nonumber \\ & -& 
S_-(k,q) [ H_e(q) +is(k)B_o(q) ]  \}
 \label{sd} 
  \\ 
0 & = &  
(s(k) \Omega - h) [B_e(k)-is(k)B_o(k) ]
\nonumber \\ & & 
-(h - s(k) \Omega+1) (H_e(k) -is(k)H_o(k) )
\nonumber \\ & + & 
\frac{2}{N}  \sum_{q}  [ (s(k) \Omega - h) (C_+(k,q) B_e(q)
 \nonumber \\ &- & 
is(k) S_+(k,q) B_o(q) )  
-(h - s(k) \Omega+1) 
 \nonumber \\ & & 
(S_+(k,q) H_e(q)
-is(k) C_+(k,q) H_o(q) ) ] 
\nonumber \\ 
 \label{su} 
\end{eqnarray}
with:
\begin{eqnarray}
C_{\pm}(k,q) & \equiv &   \int_{-\infty}^{\infty} dx \cos(qx)\cos(kx)(e^{\pm |k| \eta (x)}-1)
\nonumber \\
S_{\pm}(k,q)  & \equiv &   s(k) \int_{-\infty}^{\infty} dx \sin(qx)\sin(kx)(e^{\pm|k| \eta (x)}-1)
\nonumber \\
\end{eqnarray}
Considering the cases $s(k) = sign(k)=\pm 1$, one gets four Extinction equations from Eqns. (\ref{sd},\ref{su}):
\begin{eqnarray}
 0 & = & B_e(k) -iB_o(k) -H_e(k) 
 +iH_o(k) 
 \nonumber \\ & +&
 \frac{2}{N} \sum_{q}  \{  C_-(k,q) [ B_e(q) 
 +i H_o(q) ] \nonumber \\ & -& 
S_-(k,q) [ H_e(q) +iB_o(q) ]  \}
 \label{sdp} 
  \\ 
  0 & = & B_e(k) +iB_o(k) -H_e(k) 
 -iH_o(k) 
 \nonumber \\ & +&
 \frac{2}{N} \sum_{q}  \{  C_-(k,q) [ B_e(q) 
 -i H_o(q) ] \nonumber \\ & -& 
S_-(k,q) [ H_e(q) -iB_o(q) ]  \}
 \label{sdm} 
  \\ 
0 & = &  
( \Omega - h) [B_e(k)-iB_o(k) ]
\nonumber \\ & & 
-(h - \Omega+1) (H_e(k) -iH_o(k) )
\nonumber \\ & + & 
\frac{2}{N}  \sum_{q}  [ ( \Omega - h) (C_+(k,q) B_e(q)
 \nonumber \\ &- & 
iS_+(k,q) B_o(q) )  
-(h -  \Omega+1) 
 \nonumber \\ & & 
(S_+(k,q) H_e(q)
-i C_+(k,q) H_o(q) ) ] 
 \label{sup} \\ 
0 & = &  
(\Omega + h) [B_e(k)+iB_o(k) ]
\nonumber \\ & & 
+(h + \Omega+1) (H_e(k) +iH_o(k) )
\nonumber \\ & + & 
\frac{2}{N}  \sum_{q}  [ ( \Omega + h) (C_+(k,q) B_e(q)
 \nonumber \\ &+ & 
i S_+(k,q) B_o(q) )  
+(h +  \Omega+1) 
 \nonumber \\ & & 
(S_+(k,q) H_e(q)
+i C_+(k,q) H_o(q) ) ] 
\nonumber \\ 
 \label{sum} 
\end{eqnarray}

By defining new variables:
\begin{equation}
\begin{array}{ccc}
U_{\pm} \equiv B_e \pm i H_o & , & V_{\pm} \equiv H_e \pm i B_o
\label{UVpm}
\end{array}
\end{equation}
and using the matrices $P_{\pm}, R_{\pm}$ defined in Eqns. (\ref{pr}), Eqns. (\ref{sdp}-\ref{sum}) become:

\begin{eqnarray}
 0 & = & P_- U_+ - R_- V_+
 \label{sdpm} \\ 
  0 & = & P_- U_- - R_- V_-
  \label{sdmm} \\
  0 & = &  (\Omega-h-1/2)(P_+U_- +R_+ V_-)
  \nonumber \\ & & 
  -R_+ V_+/2+P_+U_+/2
 \label{supm}  \\
0 & = &  (\Omega+h+1/2)(P_+U_+ +R_+ V_+)
\nonumber \\ & & 
+R_+ V_-/2-P_+U_-/2
 \label{summ} 
\end{eqnarray}
In the case of no geometric defects these equations become $U_{+}=V_+$, $U_{-}=V_-$, and:
\begin{eqnarray}
  0 & = &  (\Omega-h-1/2)U_- 
 \label{snp} \\
0 & = &  (\Omega+h+1/2)U_+
 \label{snm} 
\end{eqnarray}
The modes resulting from Eq. (\ref{snp}) correspond to surface modes of positive frequency $\Omega = h+1/2$, with $U_-(k)=V_-(k) \neq 0$, $U_+(k)=V_+(k)=0$. The latter (using Eqns. (\ref{UVpm})) leads to $B_e=-iH_o$, $H_e=-iB_o$, and together with $U_-=V_-$ leads to $H_o=iH_e$, $B_o=iB_e$. The last equality leads to:
\begin{equation}
B(x) =(2/N)[B_e \cos (k x)+B_o \sin (kx)]
= (2/N)B_e e^{i k x} \; , 
\end{equation}
 meaning that there is only propagation to the right on the semi-infinite surface, as announced (here $k>0$, and the time dependence is $\exp (-i \omega t)$). 
 
 By using the following matrices, $T \equiv P_-^{-1}R_-$, $S \equiv P_+T+R_+$ and $Q \equiv P_+ T-R_+$, the system of Eqns. (\ref{sdpm}-\ref{summ}) becomes a regular eigenvalue ($\Omega$) problem:
 
\begin{equation}
0=
\left(
\begin{array}{cc}
(h+1/2)I-\Omega I & -QS^{-1}/2 \\
QS^{-1}/2 & -(h+1/2)I -\Omega I
\end{array}
\right) 
\left(
\begin{array}{c}
S V_- \\
S V_+
\end{array}
\right)
\label{efsi}
\end{equation}
A final step leads to:
\begin{equation}
0=
\left(
\begin{array}{cc}
 [(h+1/2)^2 -\Omega^2] I  & -(QS^{-1})^2/4
\end{array}
\right) 
\left(
\begin{array}{c}
S V_+
\end{array}
\right)
\label{eqsif}
\end{equation}
Thus, the eigenvalues $\Omega^2$ are equal to the eigenvalues of the matrix $-(QS^{-1})^2/4$ plus $(h+1/2)^2$.

\subsection{Case of thin film with flat surfaces: \label{flat}}

The spin wave modes of a film with flat surfaces may be solved by the standard method, i.e. by solving the magnetostatic Maxwell equations subject to boundary conditions, and the Landau Lifshitz equations (see discussion in section \ref{lsw}). The plane wave solution for the magnetostatic potential is of the form:
\begin{equation}
 \phi (x,y,t) = e^{i(kx-\omega t)}\phi (y) \; ,
 \end{equation}
with $\phi (y)$ given by:
\begin{equation}
\begin{array}{ccc}
\phi_u(y) =  U e^{-|k|(y-l)} & , & y >l  \\
\phi_i(y)  =  I_+ e^{|k|y} +I_- e^{-|k|y} 
& , & -l < y < l \\
\phi_d(y)  =  D e^{|k|(y+l)} & , & y < -l
\end{array}
\label{fiy}
\end{equation}
Notice that the profile $f(y)$ of the magnetostatic plane wave of Eqns. (\ref{sso}) corresponds to $\phi (y)$ evaluated for $s = sign(k) >0$.  Applying the boundary conditions of continuous magnetostatic potential and $y$ component of the magnetic induction $b_y$ at the surfaces $y=\pm l$ of the film, one obtains:
\begin{equation}
0=\left( 
\begin{array}{cc}
(\mu+1+s\nu)e^{|k|l} & (-\mu+1+s\nu)e^{-|k|l} \\
(\mu-1+s\nu)e^{-|k|l} & (-\mu-1+s\nu)e^{|k|l}
\end{array}
\right) 
\left( 
\begin{array}{c}
I_+ \\ I_-
\end{array}
\right)
\end{equation}
which is equivalent to 
\begin{equation}
0=\left( 
\begin{array}{cc}
(2+\frac{1}{h+s\Omega})e^{|k|l} & -\frac{1}{h-s\Omega}e^{-|k|l} \\
\frac{1}{h+s\Omega}e^{-|k|l} & -(2+\frac{1}{h-s\Omega})e^{|k|l}
\end{array}
\right) 
\left( 
\begin{array}{c}
I_+ \\ I_-
\end{array}
\right)
\label{ipm}
\end{equation}
Imposing the determinant of the previous matrix to be null leads to the frequencies of the DE surface modes of Eq. (\ref{DEf}) (there are solutions with equivalent negative frequencies). From the eigenvector associated with Eq. (\ref{ipm}) at positive frequency (fixed value of $k$) one deduces that the ratio of the potentials evaluated at the surfaces is:
\begin{equation}
\frac{\phi (l)}{\phi (-l)} =
\frac{U}{D} = \sqrt{\frac{ h+1/2-s \Omega}{
h+1/2 +s \Omega}} \; ,
\label{UD}
\end{equation}
which shows that the amplitude of the spin wave in propagation to the right ($\Omega>0$ and $k>0$) is significant in the lower surface, and vice versa for propagation to the left i.e. in that case significant in the upper surface.

 In the case of a film with flat surfaces one may also obtain the anti-symmetric and symmetric  modes of Eqns. (\ref{SA},\ref{SAi}) through our Extinction equations. Indeed, Eqns. (\ref{a3}-\ref{af2}) for the anti-symmetric modes become effectively $B_e^a=H_e^a$, $B_o^s=H_o^s$, and:
 
 \begin{eqnarray}
0 & = & -2\Omega D_+H_o^s-i[(2 h +1)D_+ +D_-] H_e^a
\nonumber \\ 
0 & = & i[(2 h+1) D_+ -D_-) H_o^s-2\Omega D_+H_e^a
\; , \label{efas} 
\end{eqnarray}
These equations may be written in matrix form as:
\begin{equation}
0=
\left(
\begin{array}{cc}
-\Omega I & -i[(h+\frac12) I+\frac{D_-^2}{2}] \\
i[ (h+\frac12) I-\frac{D_-^2}{2}] & -\Omega I
\end{array}
\right) 
\left(
\begin{array}{c}
D_+ H_o^s \\
D_+ H_e^a
\end{array}
\right)
\label{mfas}
\end{equation}

Imposing the determinant of the previous equations to be null, leads to the dispersion relation of Damon Eshbach (DE) modes. 
For the positive frequencies (there are $\pm$ pairs of solutions), the eigenvector corresponds to:
\begin{equation}
\frac{iH_o^s(k)}{H_e^a(k)}=\sqrt{\frac{h+1/2+e^{-2|k|l}/2}{h+1/2-e^{-2|k|l}/2} }\; .
\label{HA}
\end{equation}

Also, Eqns. (\ref{s3}-\ref{sf2}) for the symmetric modes become effectively $B_e^s=H_e^s$, $B_o^a=H_o^a$, and:
 
 \begin{eqnarray}
0 & = & -2\Omega D_+H_o^a-i[(2 h +1)D_+ -D_-] H_e^s
\nonumber \\ 
0 & = & i[(2 h+1) D_+ +D_-) H_o^a-2\Omega D_+H_e^s
\; , \label{efs} 
\end{eqnarray}

These equations may be written in matrix form as:
\begin{equation}
0=
\left(
\begin{array}{cc}
-\Omega I & -i[(h+\frac12) I-\frac{D_-^2}{2}] \\
i[ (h+\frac12) I+\frac{D_-^2}{2}] & -\Omega I
\end{array}
\right) 
\left(
\begin{array}{c}
D_+ H_o^a \\
D_+ H_e^s
\end{array}
\right)
\label{mfss}
\end{equation}
Imposing the determinant of the previous equations to be null, leads again to the frequency eigenvalues of the DE modes of Eq. (\ref{DEf}).
For the positive frequencies, the eigenvector corresponds to:
\begin{equation}
\frac{iH_o^a(k)}{H_e^s(k)}=\sqrt{\frac{h+1/2-e^{-2|k|l}/2}{h+1/2+e^{-2|k|l}/2} }\; .
\label{HS}
\end{equation}
The results for the eigenvectors of the anti-symmetric and symmetric modes of Eqns. (\ref{HA},\ref{HS}) may be verified by use of the “standard" solution of Eq. (\ref{fiy}), indeed more directly by use of Eq. (\ref{UD}).

\bibliographystyle{unsrt}
\bibliography{Citations.bib}

  \end{document}